\documentclass[%
reprint,
superscriptaddress,
 amsmath,amssymb,
 aps,
]{revtex4-1}

\usepackage{graphicx,setspace}% Include figure files
\usepackage{dcolumn}% Align table columns on decimal point
\usepackage{bm}% bold math
\usepackage[mathlines]{lineno}% Enable numbering of text and display math
\usepackage[colorlinks=true,citecolor=blue,linkcolor=red,anchorcolor=green,urlcolor=cyan]{hyperref}
\usepackage{mathrsfs}
\usepackage{xcolor}
\usepackage{float}
\usepackage{ulem}
\usepackage[utf8]{inputenc}

\newcommand{\be}{\begin{equation}}
\newcommand{\ee}{\end{equation}}
\newcommand{\bal}{\begin{aligned}}
\newcommand{\eal}{\end{aligned}}

\begin{document}

%\preprint{APS/123-QED}

\title{NUT charges and black hole shadows}% Force line breaks with \\

\author{Ming Zhang}
\email{mingzhang@jxnu.edu.cn}
\affiliation{Department of Physics, Jiangxi Normal University, Nanchang 330022, China}
\author{Jie Jiang}
\email{corresponding author, jiejiang@mail.bnu.edu.cn}
\affiliation{Department of Physics, Beijing Normal University, Beijing 100875, China}

\date{\today}% It is always \today, today,
 
\begin{abstract}
We study the nontrivial effects of the NUT charges on the shadows of the Kerr-Taub-NUT black holes seen by zero-angular-momentum-observers. Inclination angles with which the observers gain maximal, locally extreme or minimal shadow sizes and distortions are investigated for the black holes with different NUT charges and distributions of string singularities. Typically, we discover that when the observer approaches the string singularity, the shadow size can be relatively very small while the shadow distortion being relatively quite large. And exceptionally, we find that when the string singularity locates only at the south pole axis of the black hole with large enough NUT charge, the shadow size is maximal for a north pole observer. 

%\begin{description}
%04.70.Dy, 04.70.-s
%\end{description}
\end{abstract}

%\pacs{04.70.Dy, 04.70.-s}% PACS, the Physics and Astronomy

\maketitle

%\tableofcontents

\section{Introduction}%\label{Introduction}
Supposing there are light sources everywhere except the zone between the black hole and the observer, there will be a dark section in the sky of the observer. This  optical appearance dark sky, dubbed as black hole shadow, is formed by motions of the photons influenced by the strong gravity system they passed by in vicinity, or in other words, by the gravitational lensing effect \cite{Cunha:2015yba}. The boundary of the shadow  directly corresponds to the  apparent image of the photons revolving the black hole on the closed bound orbits  that are unstable.

Shadows or the apparent shape of a spherically symmetric black hole, like the Schwarzschild black hole case \cite{synge1966escape,luminet1979image}, is perfectly circular in view of an observer with any inclination angle. The shadow of an axially symmetric black hole, like the Kerr black hole case \cite{hawking1973black}, is optically deformed and elongated silhouette-like  in the direction of the rotating axis. Many representative investigations as extensions of the Kerr case have been done\cite{Gralla:2019xty,Hioki:2008zw,Grenzebach:2014fha,Wang:2017qhh,Guo:2018kis,Hennigar:2018hza,Konoplya:2019sns,Bambi:2010hf,Konoplya:2019fpy,Wei:2018xks,Jusufi:2020odz,Wang:2018prk,Liu:2020ola,Kumar:2019pjp,Chang:2020miq,Zeng:2020dco,Zeng:2020vsj,Xavier:2020egv,Hu:2020usx,Jusufi:2019ltj,Ghasemi-Nodehi:2020oiz,Lu:2019zxb,Feng:2019zzn,Hendi:2020ebh,Cunha:2018cof,Cunha:2018gql,Gralla:2020yvo,Gralla:2020yvo,Johnson:2019ljv,Zhang:2019glo}, also including the naked singularity case \cite{Hioki:2009na} and the wormhole cases \cite{Ohgami:2015nra,Nedkova:2013msa,Shaikh:2018kfv,Amir:2018szm,Amir:2018pcu,Wang:2020emr}. There are correspondences between characteristics including the size and the shape (usually reflected by distortion) of the apparent shape and the parameters of the background geometry. It is this correspondence that makes the observational test of the general relativity together with its modifications possible \cite{Johannsen:2016uoh,Akiyama:2019cqa}.

Almost all the shadows of the rotating  black holes that have been studied share common characteristics:  the size and the distortion take maximal values for an equatorial observer and take minimal values for a polar observer. However, there are special cases: the shadow of the black hole with acceleration that has been investigated in our recent paper \cite{Zhang:2020xub} and the black hole with gravitomagnetic mass that will be reported in the following. Both of these two kinds of black hole belongs to the Plebański-Demiański spacetime solution in the Einstein gravity \cite{Griffiths:2005qp}. In \cite{Zhang:2020xub}, we found it is an off-equatorial observer who gains maximal size and distortion for the accelerating Kerr black hole. In this paper, we would like to show our investigation of  the Kerr-Taub-NUT (KTN) black hole's shadow, presenting more nontrivial effects of the NUT charges on the size and shape of the black hole's apparent shape.

The KTN spacetime, which owns three hairs: the gravitoelectric charge, i.e., the gravitational mass; the gravitomagnetic mass, i.e., the NUT charges or the Misner gravitational  charges; the angular momentum. The spacetime is  asymptotically non-flat due to the NUT charge, and there are string singularities on the symmetric axis. In Ref. \cite{Abdujabbarov:2012bn}, the shadow of the black hole was investigated; however, the observer was merely considered on the equatorial plane, the different distributions of the string singularities were neglected  and the cases with and without ring singularities were not differentiated. In Ref. \cite{Grenzebach:2014fha}, the shadow of the black hole (with electric charge and cosmological constant) was studied; nevertheless, what the authors emphasized there is the analytical formula for the shadows seen by Carter observers at finite distance, and what they mainly found is that the shadow of the black hole with NUT charge is always symmetric to the horizontal axis, even for an off-equatorial observer. So, questions remain there: firstly, of observers on equator plane, off-equator plane and pole, which one can see the maximal or minimal shadow size and distortion? secondly, how does the distribution of the string singularities affect the apparent shape? thirdly, what are the differences of the shadows of the black hole with or without central singularity? 

We in this paper are to answer these questions. In the next section, we will derive an analytical formula of the shadow of the KTN black hole for a zero-angular-momentum-observer (ZAMO). In Sec. \ref{sec4}, we will analyze the observables of the shadow, and Sec. \ref{sec5} is devoted to our conclusions. We note that in this paper we will consider the appearance of the KTN black hole illuminated by a uniform, isotropically emitting screen (surrounding the black hole) being distant and spherical, which is pedagogically useful to deepen our understanding of the lensing features of the KTN black hole and inspiring to investigate other (more realistic) cases where the black hole is illuminated by  a planar screen, or the photons are emitted from optically and geometrically accretion  disks which could be either thin or thick \cite{Gralla:2019xty}. And in the terminology of this paper, the ``shadow" corresponds to the interior of the critical curves \cite{Gralla:2019xty} or circle of apparent boundary \cite{bardeen1972jm}.

\section{Shadow Contours of the KTN black holes}\label{sec2}
The line element of the KTN black hole is \cite{Griffiths:2005qp}
\be\label{metric}
\bal
ds^2=&\Sigma  \left(d\theta^2+\frac{dr^2}{\Delta _r}\right)-\frac{ \left(\Delta _r-a^2\sin ^2\theta \right)dt^2}{\Sigma }\\ &+\frac{2  \left[\chi  \Delta _r-a \sin ^2\theta (a \chi +\Sigma )\right] dt d\phi}{\Sigma }\\&+\frac{ \left[  (a \chi +\Sigma )^2 \sin ^2\theta -\chi ^2 \Delta _r\right] d\phi^2}{\Sigma },
\eal
\ee
where
\be\bal
\chi&=a\sin ^2\theta-2l (\cos\theta+C),\\
\Sigma&=r^{2}+(l+a \cos\theta )^2,\\
\Delta_r&=r^2-2mr+a^2-l^2,\\ \nonumber
\eal\ee
$m,\, a,\,l$ are black hole parameters for mass, rotation and NUT charge. The Manko–Ruiz parameter $C$ \cite{Manko:2005nm} is introduced to adjust the singularity on the rotation axis (or the position of the Misner string) brought by the NUT charge and it is not a pure gauge parameter as the asymptotic behavior of the spacetime can be changed by it. The string singularity locates only on the north pole axis $\theta=0$ for $C=1$ and only on the south pole axis $\theta=\pi$ for $C=-1$. Misner string tubes surrounding the string singularities should be introduced  as boundaries of the spacetime because $(\nabla t)^2$ becomes ill-defined on the axis \cite{Misner:1963fr}.   The event horizon of the black hole is $r_\pm=m\pm\sqrt{m^2 -a^2+l^2}.$ If $l^2<a^2$, the spacetime curvature ring singularity is formed at $r=0$ and $\cos\theta=-l/a$; for $l^2>a^2$, the spacetime is free of ring singularity and is regular at $r=0$. Though for any value of $C$ the spacetime keeps geodesically complete, the condition of the absence of closed timelike and null geodesics that violate the causality requires $|C|\leqslant 1$ \cite{Clement:2015cxa,Bordo:2019rhu}. Besides, if $g_{\phi\phi}<0$, closed time-like $\phi$ lines emerge in this spacetime.

In the KTN spacetime, the equations of motion for the geodesic photon are  \cite{Grenzebach:2014fha}
\begin{equation}
\Sigma\frac{dt}{d\tau}= \frac{\chi(L_{z}-E\chi)}{\sin^{2}\theta}
+\frac{(\Sigma + a\chi) \left[\Sigma + a\chi)E - aL_{z}\right]}{\Delta_{r}},
\end{equation}
\begin{equation}
\Sigma\frac{d\phi}{d\tau}= \frac{L_{z}-E\chi}{ \sin^{2}\theta}
		+\frac{a\left[\Sigma + a\chi)E - aL_{z}\right]}{\Delta_{r}},
\end{equation}
\begin{equation}
\Sigma^2\left( \frac{d\theta}{d\tau}\right)^2 = K - \frac{(\chi E - L_{z})^{2}}{\sin^{2}\theta} \equiv \Theta(\theta),\label{loeff}
\end{equation}
\begin{equation}
\Sigma^2\left( \frac{dr}{d\tau}\right)^2 =\left[(\Sigma + a\chi)E-aL_{z}\right]^{2} - \Delta_{r}K\equiv R(r).\label{reff}
\end{equation}
The proper time is $\tau$, the conserved energy $E$ and angular momentum $L_z$ originate from the Killing vectors $\partial_t$ and $\partial_\varphi$. $K$ is the Carter constant \cite{Carter:1968rr}.

The motion of the photon must satisfy $R(r)\geqslant 0$ and $\Theta (\theta)\geqslant 0$. To obtain the region where photons are filled, we need
\begin{equation}
R(r_p)=0,\quad R'(r_p)=0,\quad  R''(r_p)>0,\label{co1}
\end{equation}
\begin{equation}
\Theta(\theta_p)=0,\quad \dot{\Theta}(\theta_p)=0,\quad \ddot\Theta(\theta_p)<0,\label{co2}
\end{equation}
where the radius and the latitude of the photon are denoted as $r_p$ and $\theta_p$, respectively. According to Eq. (\ref{co1}), we have 
\begin{equation}
\bar{K} = \frac{16r^{2}\Delta_{r}}{(\Delta_{r}')^{2}}=\frac{1}{a}\left( \Sigma + a\chi - \frac{4r\Delta_{r}}{\Delta_{r}’}\right)
\end{equation}
and
\begin{equation}
\bar{L} =\frac{1}{a}\left( \Sigma + a\chi - \frac{4r\Delta_{r}}{\Delta_{r}’}\right).
\end{equation}
Likewise, Eq. (\ref{co2}) yields
\begin{equation}
\bar{K} = \frac{(\chi  - \bar{L})^{2}}{\sin^{2}\theta} ,
\end{equation}
\begin{equation}
\bar{L} = \frac{\chi \Delta_\theta^{'}-2a \Delta_\theta\sin(2\theta)}{\Delta_\theta^{'}}.
\end{equation}
We have rescaled the constant $K$ and $L_z$ by $E$, as $\bar{K} =K/E^2,\quad \bar{L}=L_z/E.$

\begin{figure}[htbp!]
\begin{center}
\includegraphics[width=45mm,angle=0]{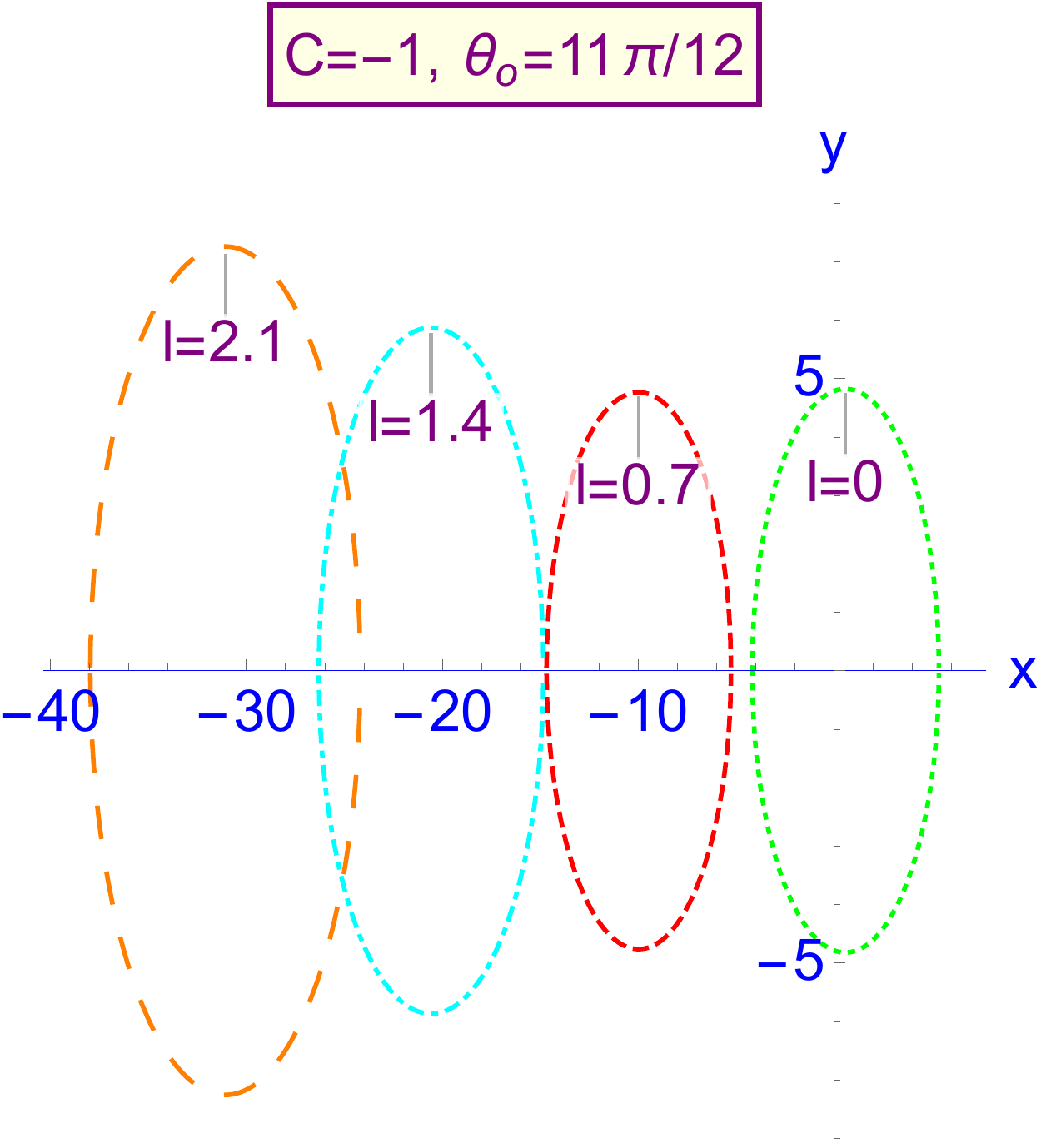}\\
\includegraphics[width=45mm,angle=0]{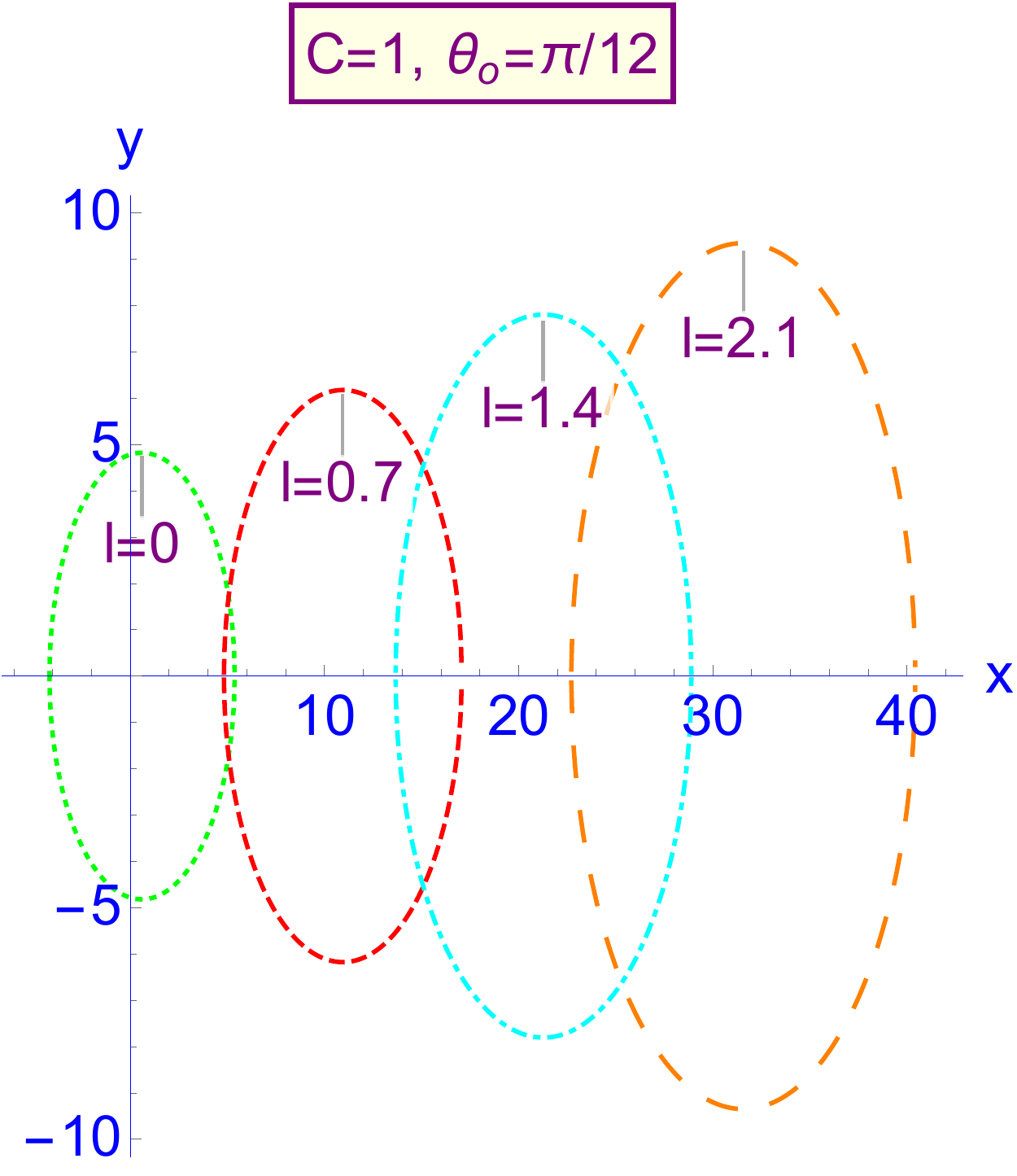}
\end{center}
\vspace{-5mm}
 \caption {Typical shadow contours of the KTN black holes with  $r_{O}=100,\,m=1\,,a=0.99$.}\label{f1}
\end{figure}

The circular orbit with constant radius and constant latitude angle for the photon cannot exist on the equatorial plane of the black hole,  due to the presence of the NUT charge \cite{Cebeci:2015fie,Guo:2020qwk}. This may bring us with different shadow contour comparing with those of the black hole owning equatorial circular orbit for the photon. To calculate the shadow contour of the KTN black hole seen by a static observer located at finite distance, we use the ZAMO reference frame
\begin{eqnarray}\label{tetrads}
  \hat{e}_{(t)} & = & \sqrt{\frac{g_{\phi \phi}}{g_{t \phi}^2 - g_{t t} g_{\phi
  \phi}}} \left( \partial_t - \frac{g_{t \phi}}{g_{\phi \phi}} \partial_{\phi}
  \right),\label{et}\\
  \hat{e}_{(r)} & = & \frac{1}{\sqrt{g_{r r}}} \partial_r,\\
  \hat{e}_{(\theta)} & = & \frac{1}{\sqrt{g_{\theta \theta}}} \partial_{\theta},\\
  \hat{e}_{(\phi)} & = & \frac{1}{\sqrt{g_{\phi \phi}}} \partial_{\phi}.
\end{eqnarray}
 Then the four-momentum of the photon with respect to the ZAMO observer can be calculated as
\begin{eqnarray} \label{xxx}
  p^{(t)} & = & - p_{\mu} \hat{e}_{(t)}^{\mu}, \\
  p^{(i)} & = & p_{\mu} \hat{e}_{(i)}^{\mu},
\end{eqnarray}
where $i = r, \theta, \phi$. Whereby the observation angles $(\alpha, \beta)$ can be defined by \cite{Cunha:2016bpi}
\begin{eqnarray}
  p^{(r)} & = & p^{(t)}\cos \alpha \cos \beta, \\
  p^{(\theta)} & = & p^{(t)} \sin \alpha, \label{ptheta} \\
  p^{(\phi)} & = & p^{(t)} \cos \alpha \sin \beta.
\end{eqnarray}
Thus we obtain the observation angles for an observer at radial position $r_O$ and inclination angle $\theta_O$ as
\be\label{sinaa}
\bal
\sin\alpha&=\frac{p^{(\theta)}}{p^{(t)}}\\&=\left.\pm\frac{1}{\zeta-\bar{L} \gamma}
  \sqrt{\frac{\Delta_\theta \bar{K}\sin^2\theta-(\chi-\bar{L})^2}{\Sigma \sin^2\theta}}\right|_{(r_O,\theta_O)},
\eal
\ee
\be
\bal
& \tan \beta = \frac{p^{(\phi)}}{p^{(r)}}\\&=\left.\frac{\bar{L} \sqrt{\Sigma \Delta_r}}
    {\sqrt{g_{\phi\phi}}\sqrt{((\Sigma+a^2\sin^2\theta)-a\bar{L})^2-\Delta_r\bar{K}}}\right|_{(r_O,\theta_O)}.
\eal
\ee
We have denoted $\zeta\equiv \hat{e}_{(t)}^t$ , $\gamma\equiv\hat{e}_{(t)}^\phi$, which are evaluated at $r=r_p$. On the plane of observer's  sky, we can define the Cartesian coordinate 
 \begin{equation}
  x \equiv - r_O \cos\alpha \sin\beta, \quad y \equiv r_O \sin\alpha .\label{xy}
\end{equation}

\begin{figure}[htbp!]
\begin{center}
\includegraphics[width=45mm,angle=0]{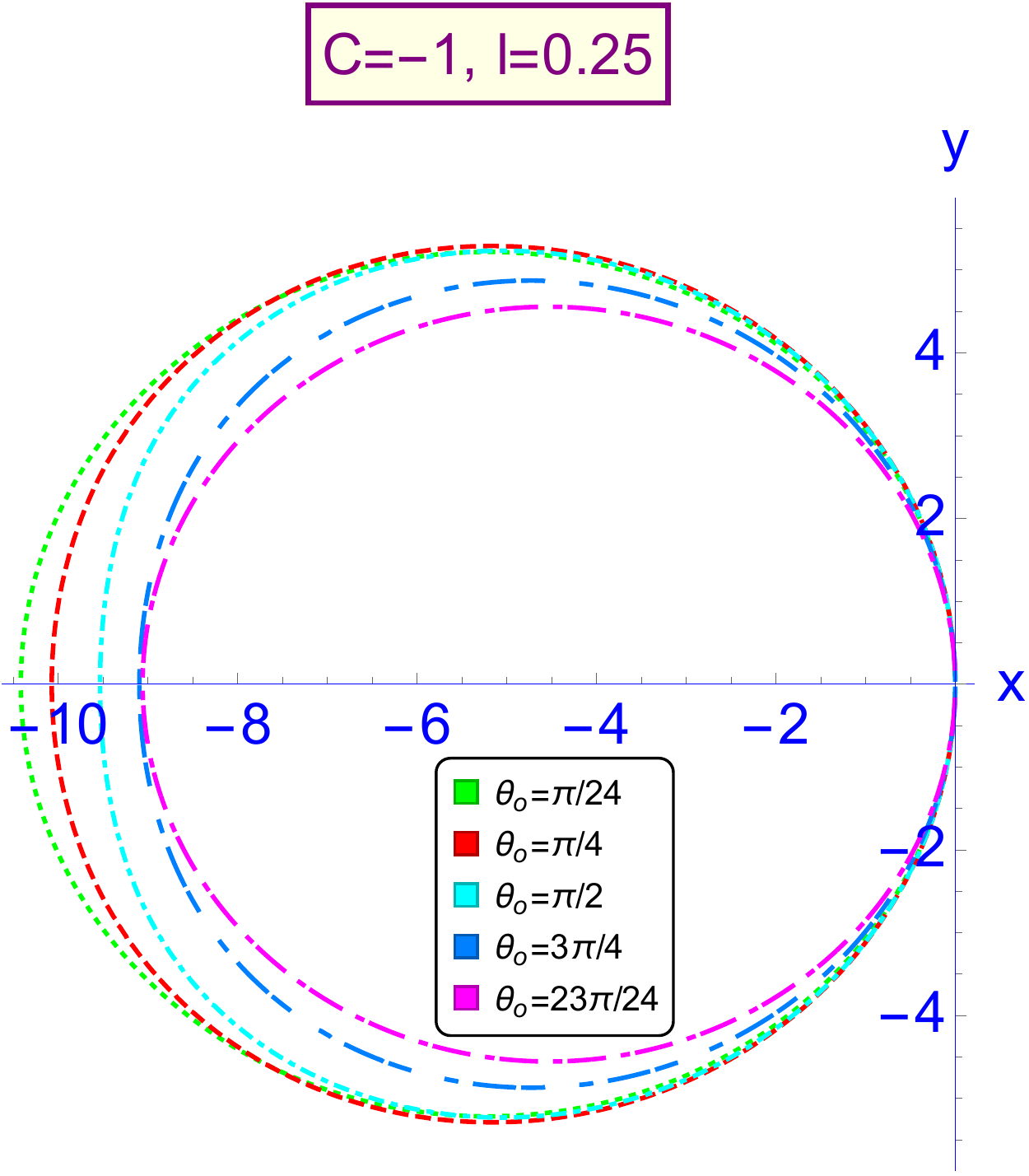}\\
\includegraphics[width=45mm,angle=0]{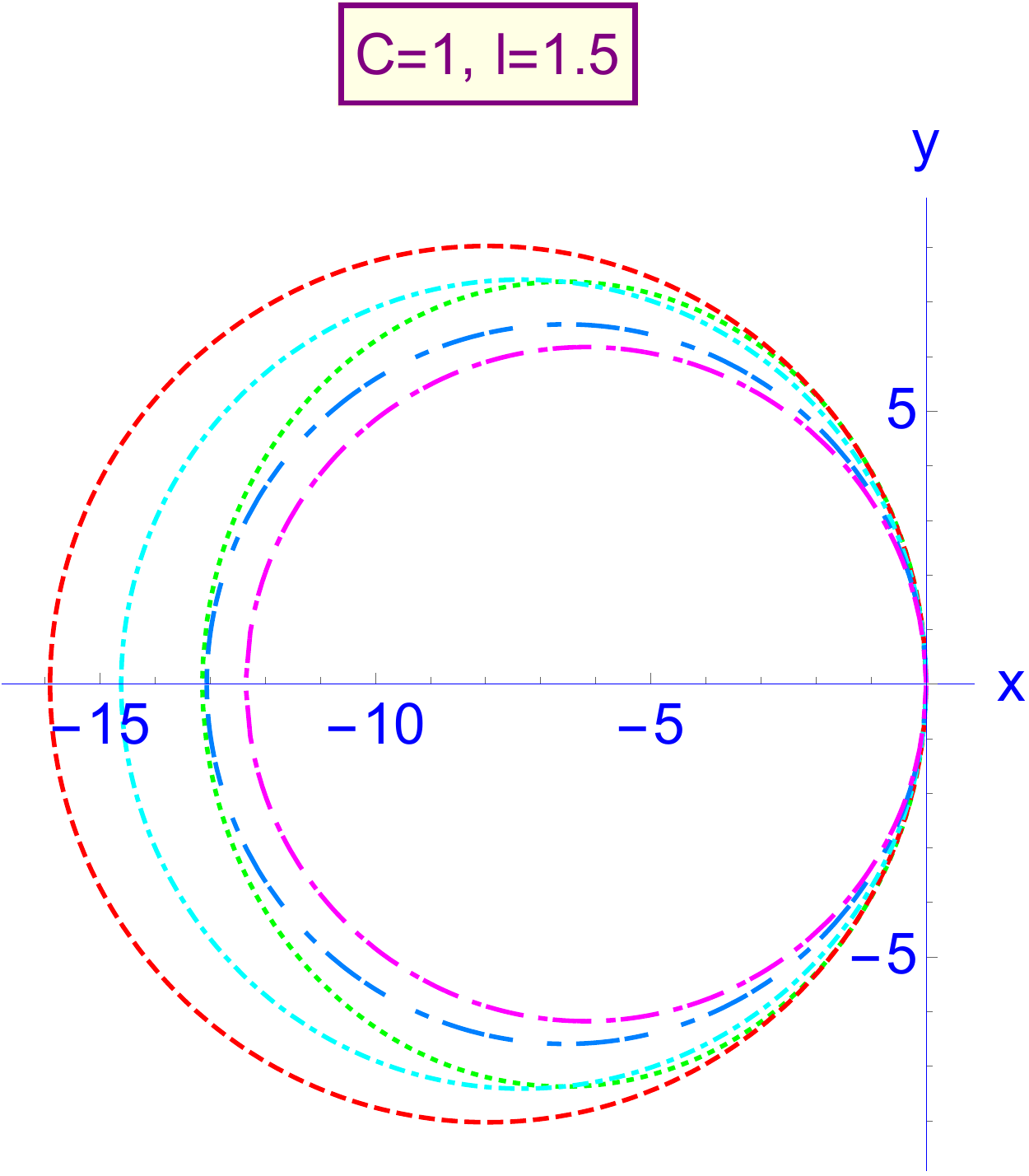}
\end{center}
\vspace{-5mm}
 \caption {Typical shadow contours of the KTN black holes with $r_O=100\,,m=1\,,a=0.99$. The rightmost points are made almost coincident manually for convenience of readers to compare the size of the shadows. }\label{fx1}
\end{figure}

The shadow contours are shown in Fig. \ref{f1}. The shadow is symmetric with respect to the horizontal axis $y=0$, as the observation angles $\alpha$ and $\pi-\alpha$ are contributed by identical reduced conserved angular momentum $\bar{L}$ and conserved Carter constant $\bar{K}$. In Fig. \ref{fx1}, shown are two typical shadow contours of the KTN black holes seen by observers with different observation angles.  In the left panel, we see that the shadow size monotonically decreases with respect to $\theta_O$ varying from $0$ to $\pi$; however, in the right panel, we know that the shadow size firstly increases but then decreases with increasing $\theta_O$. It tells us that the characteristics of the shadow for the KTN black hole are dependent on the observational angles. The changing pattern in the left panel is very strange, as usually  the size of (almost all) the black hole without NUT charge becomes largest when the observational angles are $\theta_O=\pi/2$. Even for the pattern shown in the right panel, we are not clear that which observational angle does the maximum shadow size corresponds to and it must not be $\theta_O=\pi/2$ as the circular orbit of the photon deviates from the equatorial plane due to the nonzero NUT charge. So to investigate the effect of the NUT charge on the shadow for the KTN black hole, which was not truly done in former literature, is interesting and necessary.

\section{Observables of the KTN black holes}\label{sec4}
\begin{figure}[htbp!]
\begin{center}
\includegraphics[width=45mm,angle=0]{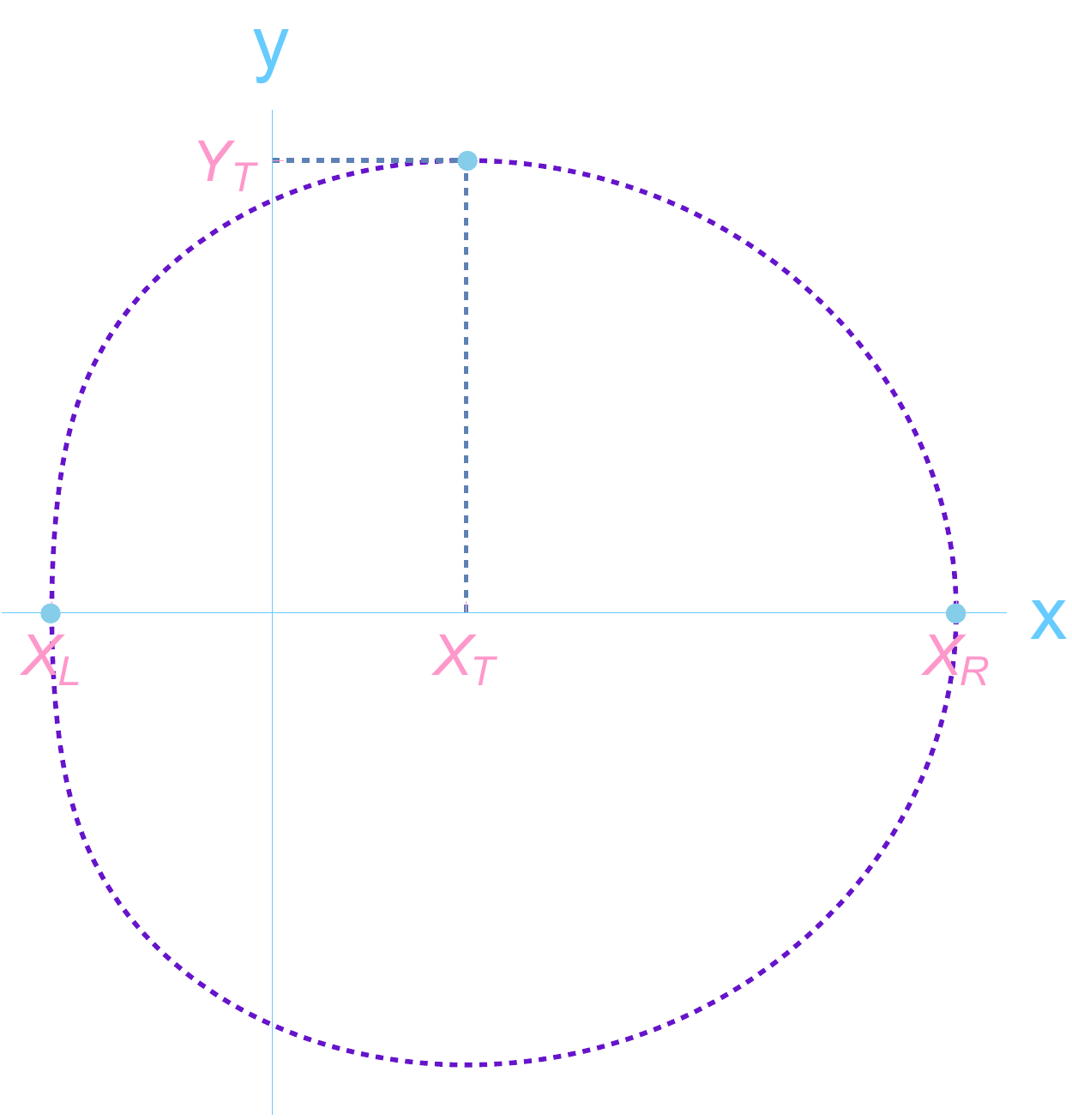}
\end{center}
\vspace{-5mm}
 \caption {Schematic diagram of the KTN black hole shadow.  $(X_{R}, 0),\, (X_{L}, 0),\,(X_{T}, Y_{T})$ correspond to the rightmost, leftmost, topmost points on the boundary, respectively.}\label{f0}
\end{figure}

We put our emphasis on studying the effects of the NUT charge on two observables of the black hole shadow: the shadow radius $R_s$ and the shadow distortion $\delta_s$. The former is the radius of an imaginary reference circle relative to the contour of the shadow; the latter reflects the shape distortion of the shadow contour comparing with the standard reference circle \cite{Hioki:2009na}. According to the schematic diagram presented in Fig. \ref{f0}, the two observables are given by \cite{Hioki:2009na}
\begin{equation}\label{defrs}
R_{s}=\frac{(X_{R}-X_{T})^{2}+Y_{T}^{2}}{2(X_{R}-X_{T})},
\end{equation}
\begin{equation}\label{defrs}
\delta_{s}=\frac{X_L-X_R+2R_s}{R_{s}},
\end{equation}
where $X_R$ and $X_L$ are the coordinate values of the rightmost and leftmost points on the shadow boundary, and $(X_T\,,Y_T)$ corresponds to the topmost point on the critical curve.

\begin{figure*}[htbp!]
\begin{center}
\includegraphics[width=55mm,angle=0]{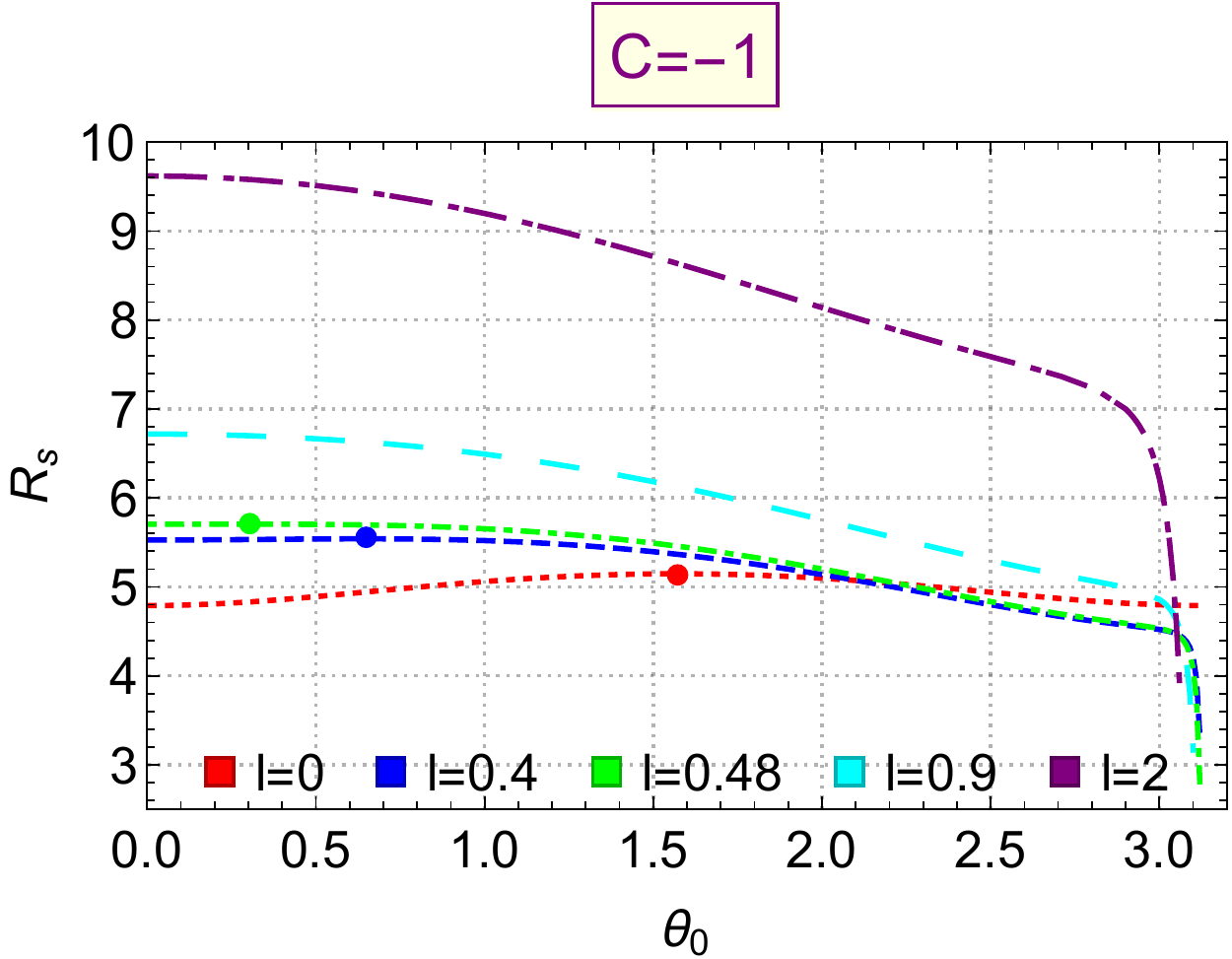}
\includegraphics[width=55mm,angle=0]{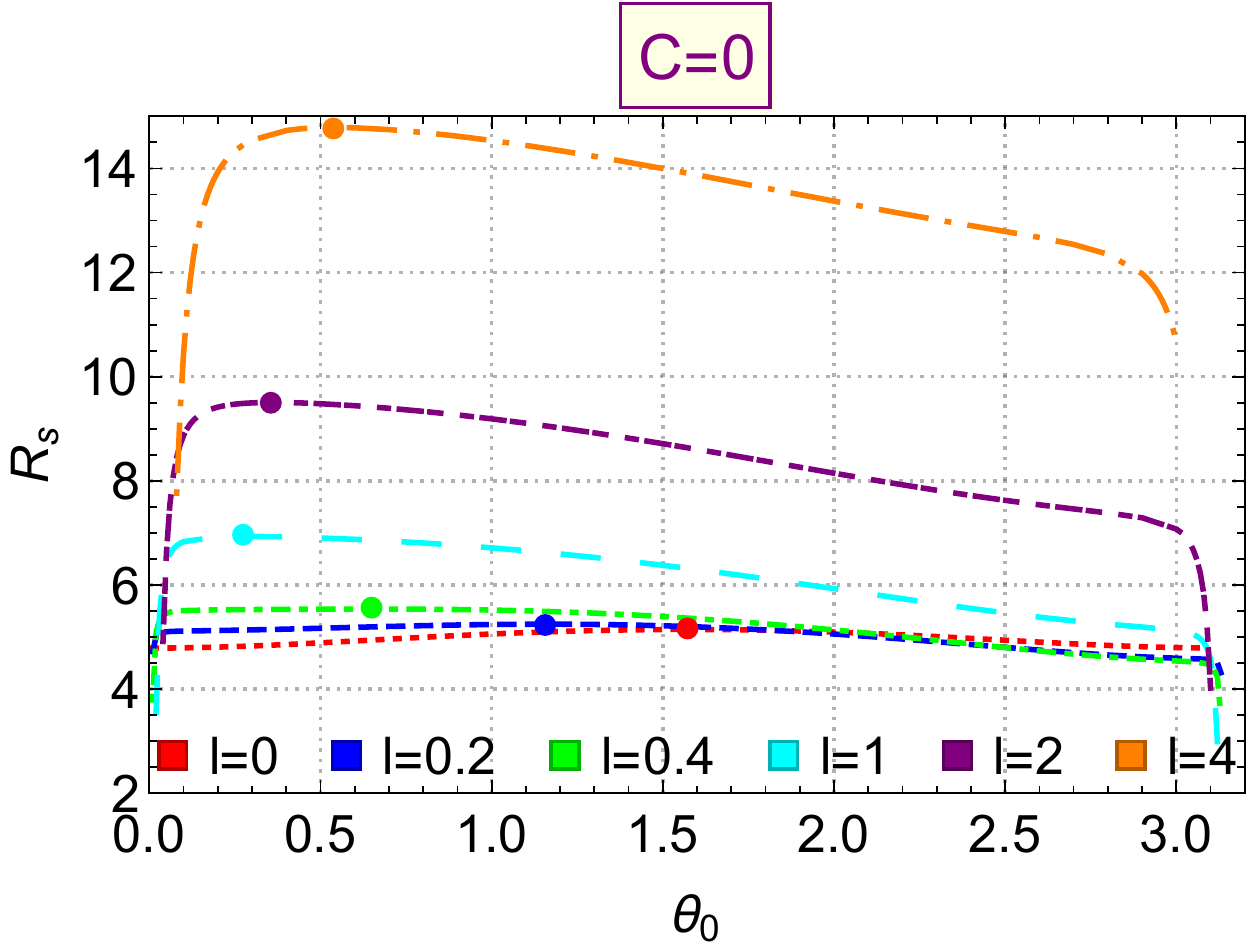}
\includegraphics[width=55mm,angle=0]{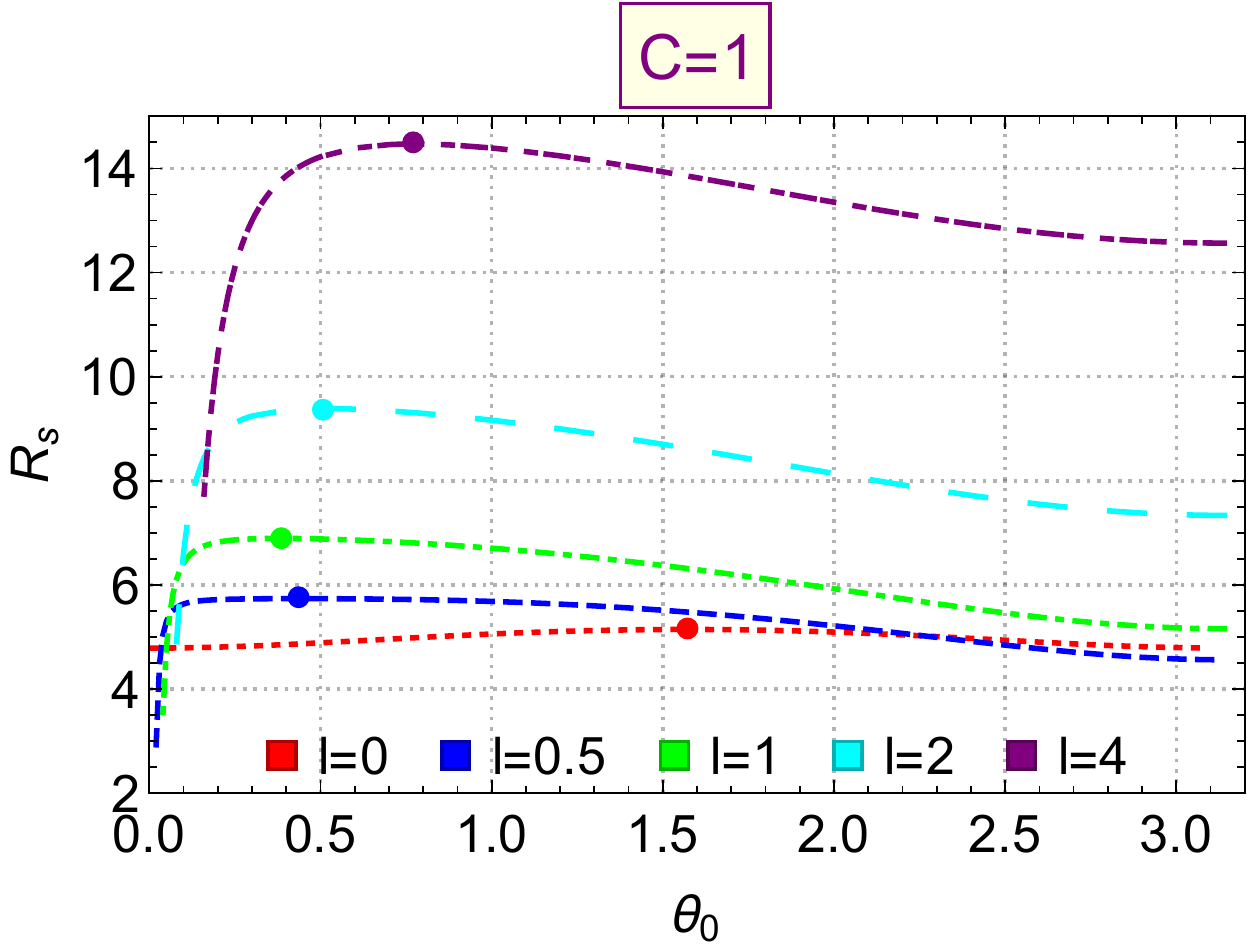}\\
\includegraphics[width=55mm,angle=0]{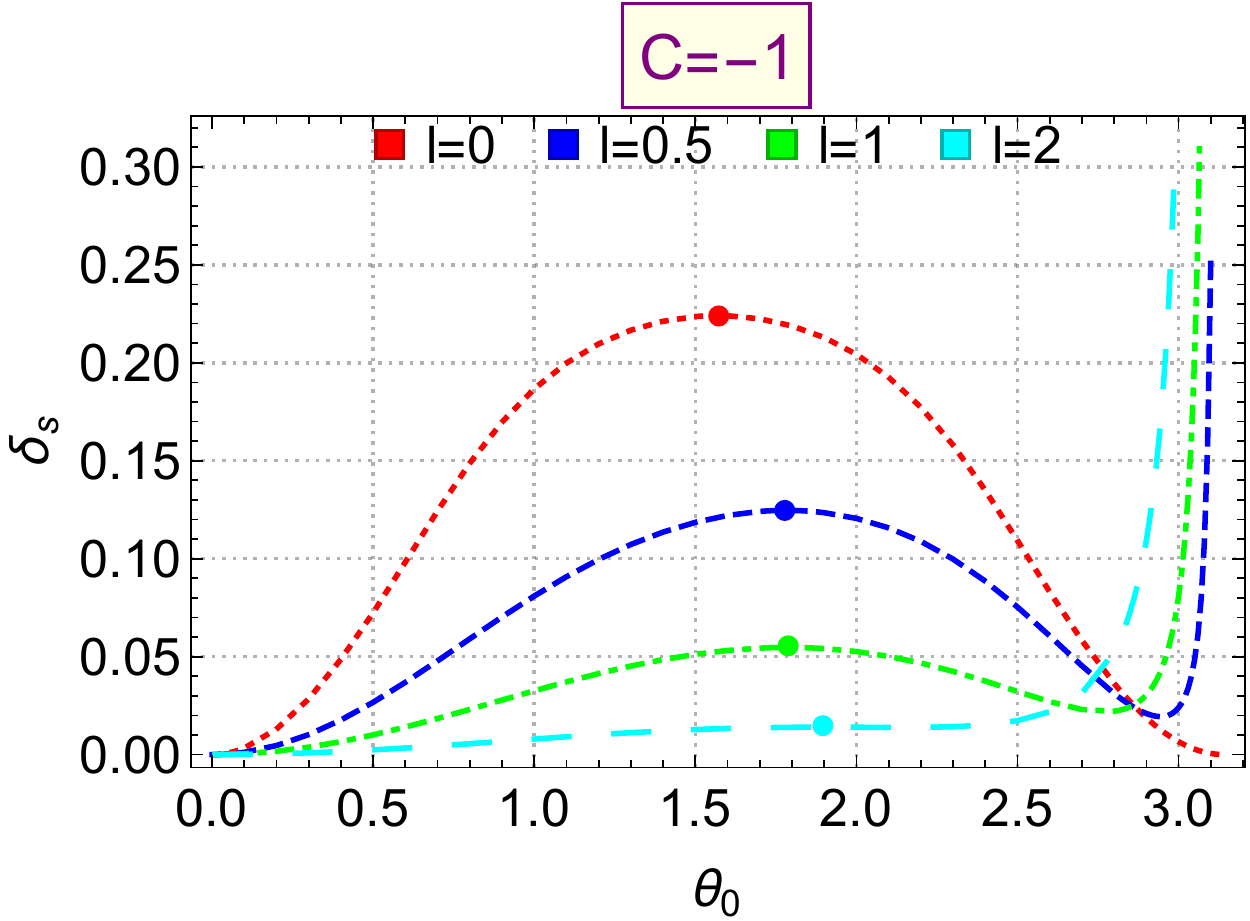}
\includegraphics[width=54mm,angle=0]{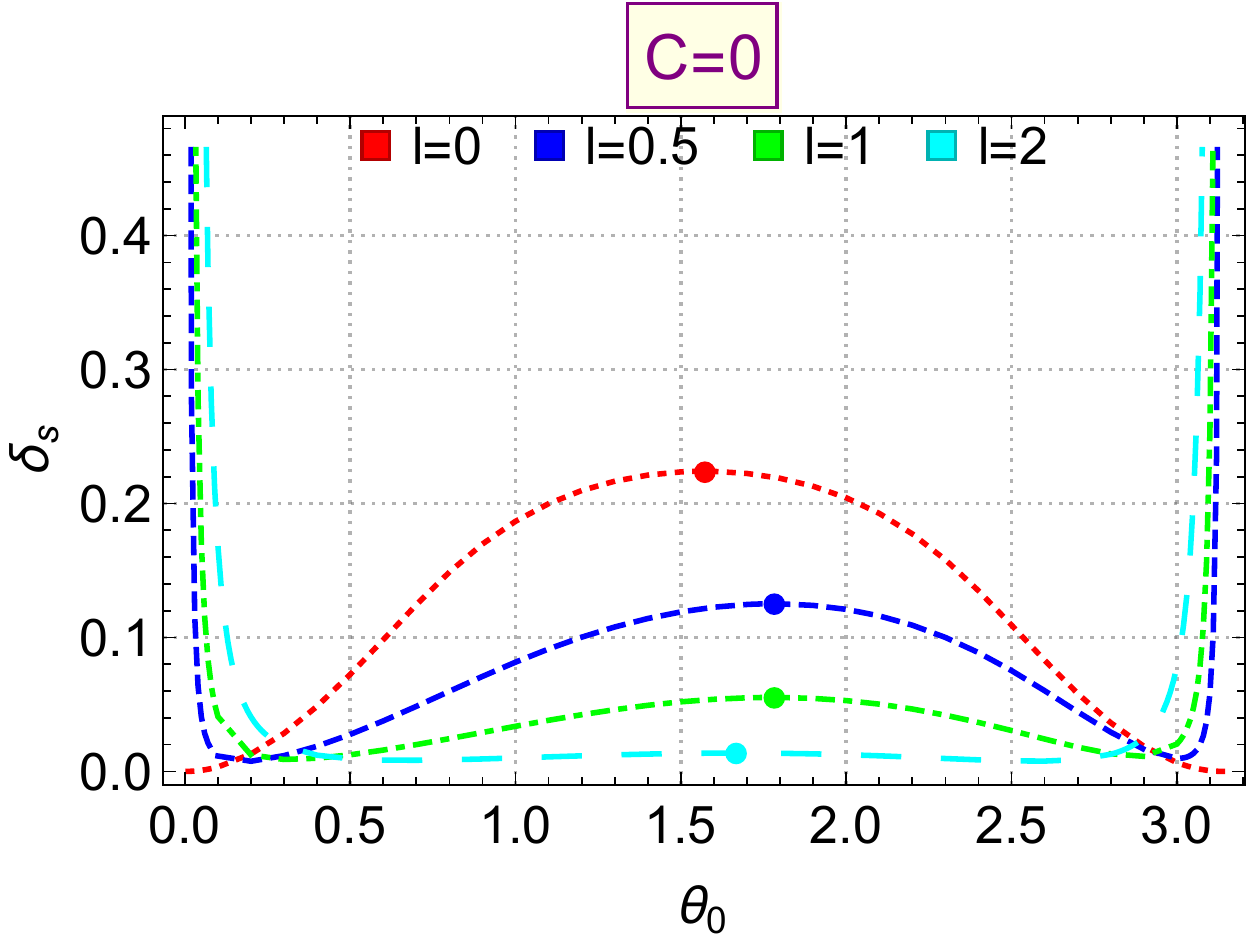}
\includegraphics[width=55mm,angle=0]{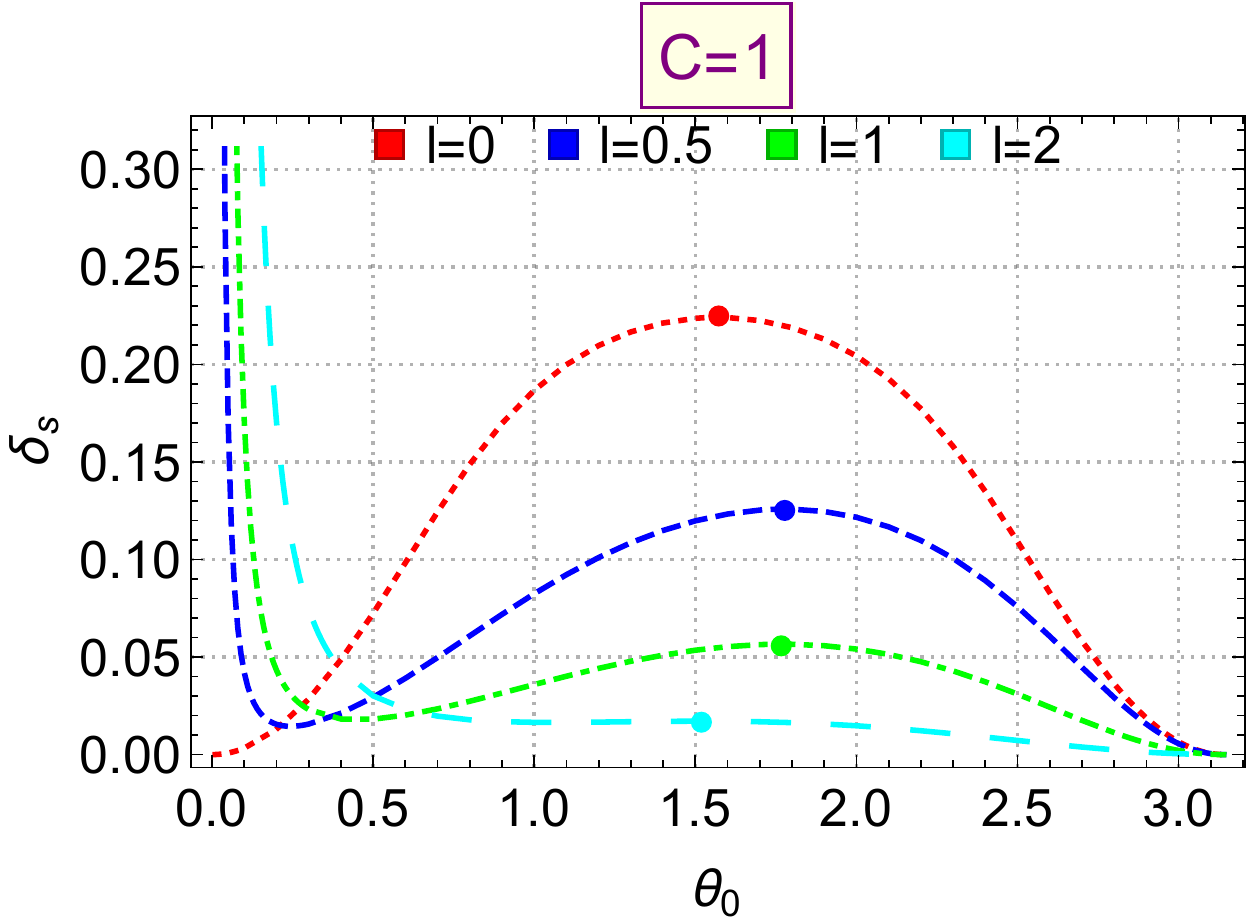}
\end{center}
\vspace{-5mm}
 \caption {The variations of the observables for the KTN black hole shadows with respect to the inclination angles of the observers with $r_O=100\,,m=1\,,a=0.99$. Top panel: the variations of the shadow sizes for $C=-1,\, 0,\,1$. Bottom panel: the variations of the shadow distortion for $C=-1,\, 0,\,1$. The dots in the diagrams denote the locally extreme value of the observables.}\label{fx}
\end{figure*}

\begin{figure*}[htbp!]
\begin{center}
\includegraphics[width=60mm,angle=0]{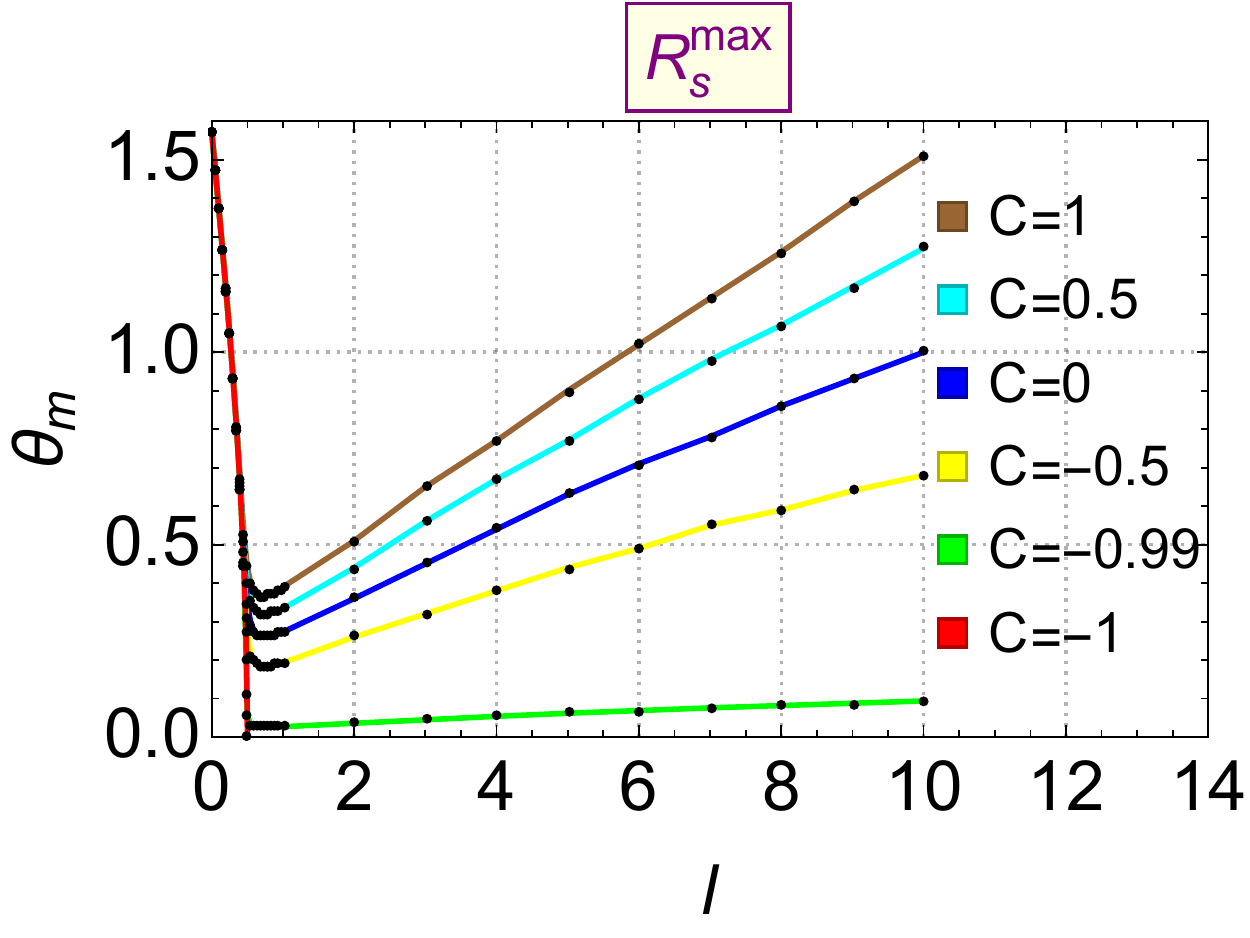}
\includegraphics[width=60mm,angle=0]{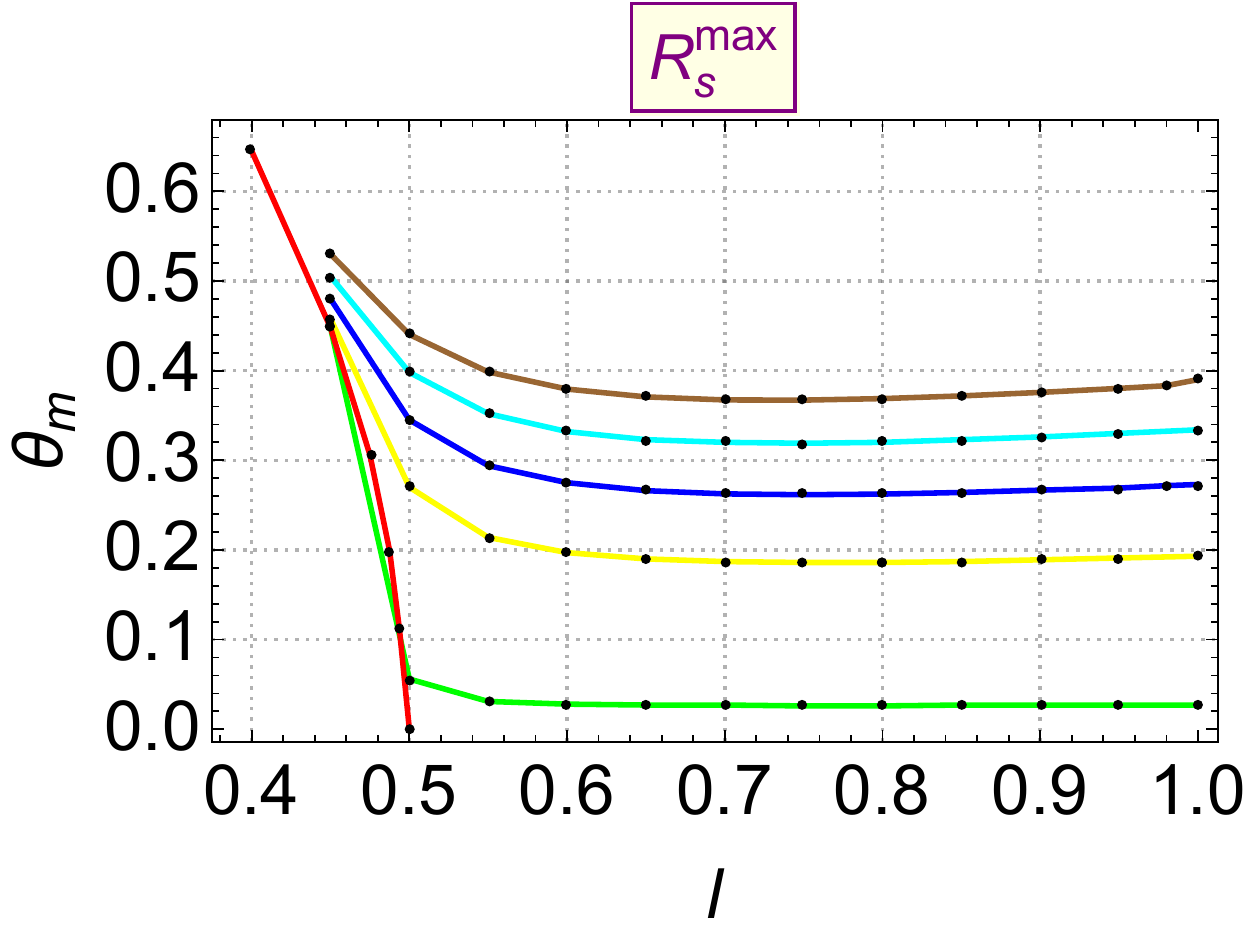}\\
\includegraphics[width=58mm,angle=0]{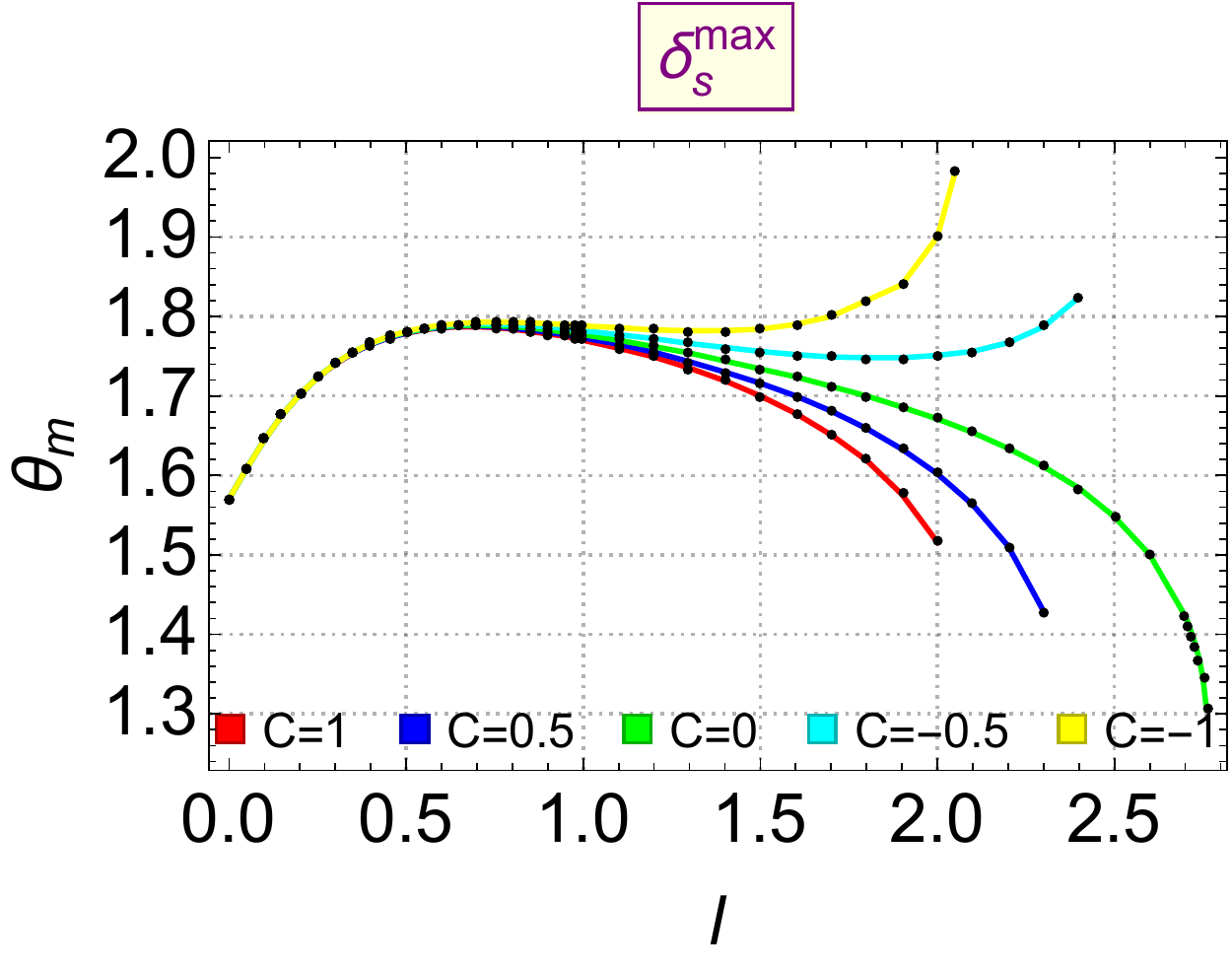}
\includegraphics[width=60mm,angle=0]{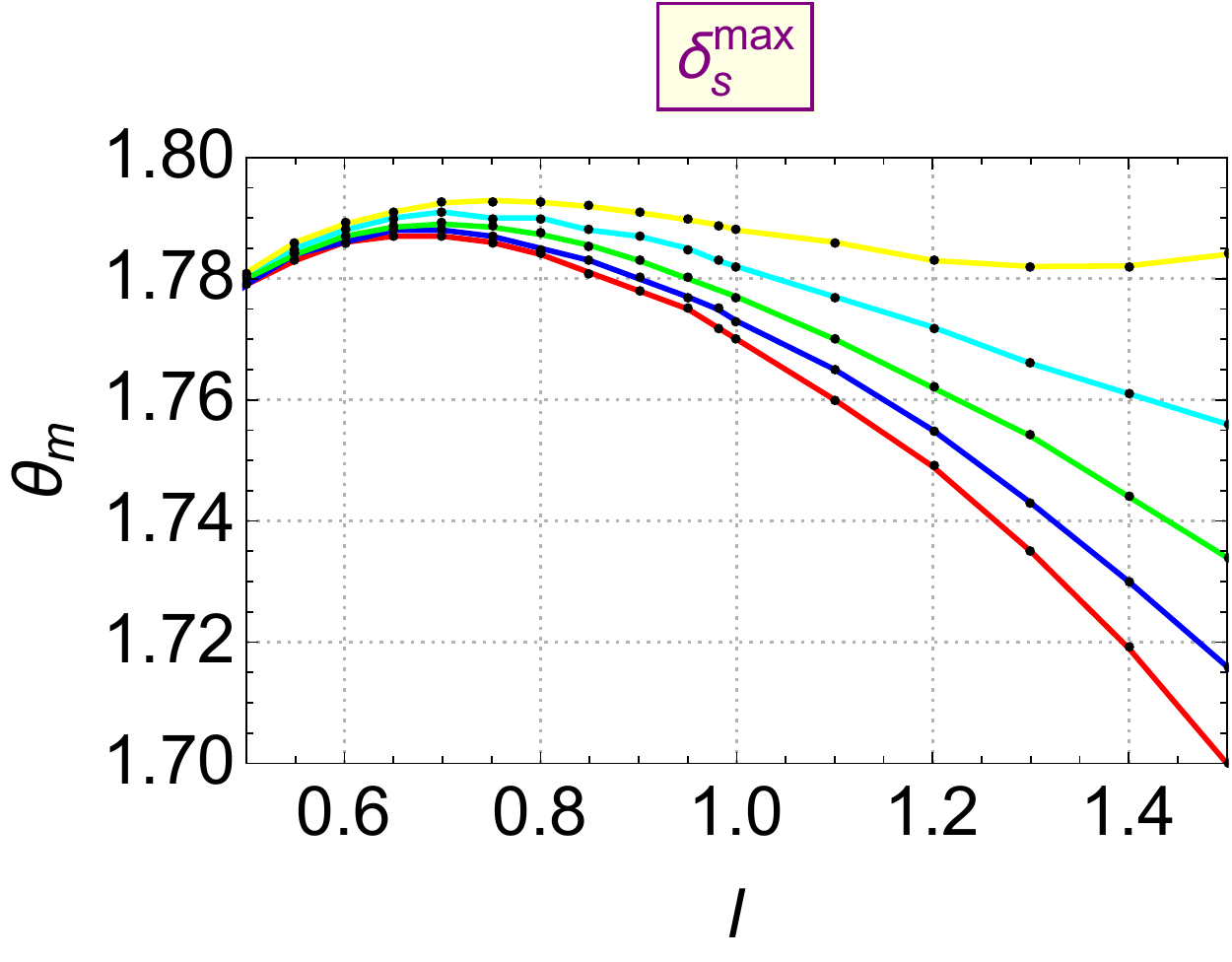}
\end{center}
\vspace{-5mm}
 \caption {Variations of the observer's inclination angles that make the shadow sizes (top panel) and the shadow distortions (bottom panel) locally maximum (as shown by the dots in Fig. \ref{fx}, we here denote them as $R_s^{\text{max}}$ and $\delta_s^{\text{max}}$) with respect to the NUT parameter for $r_O=100\,,m=1\,,a=0.99$.}
 \label{fxx1}
\end{figure*}

It is not difficult to check that $\delta_s=0$ for the Schwarzschild black holes, for which $X_T=0,\,X_R+X_L=0$. For the Kerr black hole with vanishing NUT charge $l$ and vanishing Manko–Ruiz parameter $C$, we know that the shadow radius $R_s$ and the shadow distortion $\delta_s$ becomes maximum for an equatorial observer, as shown in Figs. \ref{fx} and \ref{fxx1}. In what follows, we illustrate how does the NUT charge together with the Manko–Ruiz parameter $C$ influences the observables of the shadow based on the results obtained in Figs. \ref{fx} and \ref{fxx1}. 

{\textbf{Case $\bf{C=-1}.$}} In this case, as we mentioned in the former section, the string singularity is on the south pole axis $\theta=\pi$ and the north pole axis $\theta=0$ is regular. There are two kinds of changing styles for the shadow radius with respect to the inclination angle $\theta_O$ of the observer while only one for the shadow distortion.

For the shadow radius, type one is that the shadow radius increases first and then decreases with respect to the inclination angle $\theta_O$ varying from the north pole to the south pole for $l<l_m$ with $l_m$ some specific value less than the angular momentum of the black hole and being dependent on the specific spacetime parameters. The minimum value is obtained nearby the southern Misner string tube (not exactly at the south pole axis, as $g_{\phi\phi}<0$ and the closed time-like curve emerges there, see, for instance, Ref. \cite{Long:2018tij}); the value of the inclination angle of the observer that makes the shadow radius locally extreme decreases with respect to the increasing NUT charge $l<l_m$ . The other type of variation of the shadow radius is that the shadow radius decreases monotonically with respect to the increasing inclination angle of the observer for $l\geqslant l_m$ and the maximum value of the shadow radius is obtained by the observer only exactly at the regular north pole axis.

The shadow distortion increases drastically at the place nearby the southern Misner string tube, to a value that is greater than the locally extreme value obtained by the observer nearby the equatorial plane.  If $l<a$ where there is a ring singularity for the black hole, the value of the inclination angle of the observer that makes the distortion maximum first increases and then decreases with respect to the increasing NUT charge $l$; however, it varies in contrary tendency if $l>a$ where there is no ring singularity for the black hole.

{\textbf{Case $\bf{C=1}.$}} In this case, the string singularity is on the north pole  axis where $\theta=0$. We can see that the shadow radius drastically decreases but the distortion increases fast nearby the northern Misner string tube. 

If $l<a$, the inclination angle of the observer that makes the shadow radius maximum firstly decreases fast but then increases slowly, and the angle that makes the distortion locally extreme first increases and then decreases.

If $l> a$, the inclination angle of the observer that makes the shadow radius maximum increases monotonically with respect to the increasing NUT charge $l$, but the angle that makes the distortion locally extreme decreases monotonically when $l$ increases.

{\textbf{Case $\bf{C=0}.$}} In this case, there are string singularities on both the pole axes of the black hole. The shadow radius drastically decreases at both places nearby the two pole axes and the minimal values of the radius vary with different NUT charges $l$; The shadow distortion drastically increases at both places nearby the two pole axes and the maximal values of the distortion also vary with the NUT charge $l$. The changing tendencies of the locally extreme values of both the shadow size and the shadow distortion are just similar to the ones in the $C=1$ case.

{\textbf{Case else.}} In this case, $|C|<1$ and $C\neq 0$. We can see that, just like the ones in the $C=0$ case, the shadow radius drastically decreases and the shadow distortion drastically increases at the places nearby the two pole axes. The changing tendency of the locally extreme value of the shadow radius with respect to the varying NUT charge just behaves like the one in the $C=0$ case. When the NUT charge is fixed, the locally extreme value increases with the increasing Manko–Ruiz parameter $C$. However, for the behavior of the locally extreme value of the shadow distortion, there exists a critical value of the parameter $C$, beyond which the changing style jumps from one to the other: as we can see in the bottom left diagram of Fig. \ref{fxx1}, the $C=0.5$ case behaves like the $C=1$ one but the  $C=-0.5$ case behaves like the $C=-1$ one. Denoting the critical value of $C$ as $C_e$, we have $-1<C_e<0$, and the specific value of it depends on specific spacetime parameters.

\section{Conclusions}\label{sec5}

We investigated the  shadow of the  KTN black hole in view of a distant ZAMO observer. Due to the NUT charge, properties of the shadow observables, specifically the shadow size and distortion, change qualitatively. Without NUT charge, an equatorial observer can see a shadow of the black hole with maximum size and distortion, and a polar observer sees a shadow with minimal size and distortion; however, with the NUT charge, things do change.

We found that the distributions of the string singularities affect the maximal and minimal shadow size and distortion of the black hole greatly. If there is string singularity on the pole, an observer with an inclination angle approaching the pole will see a shadow with relatively small size and large distortion.

We explored how the NUT charge affects the locally extreme observation angles where observers get locally extreme shadow size and distortion. A striking fact is that there exists a critical NUT charge, beyond which the shadow size becomes maximal only for a northern polar observer when the string singularity locates on the south pole.

Our novel results demonstrate the extraordinary observational effect of the NUT charges and string singularities on the shadows of the KTN black holes. It reveals that for the Kerr or Kerr-like black holes with NUT charges, the common property of the shadow sizes and distortions being minimal for polar observers and maximal for  equatorial observers will not be valid any more. Nevertheless, it should be noted that the shadows of black holes deviating considerably from the Kerr and without NUT charges \cite{Cunha:2019dwb,Herdeiro:2020wei} may also display remarkable differences  in some cases for equatorial  plane observations (cf. Refs. \cite{Cunha:2015yba,Cunha:2016bjh} for shadows of non-Kerr-like black holes with boson star-like hairs).

\section*{Acknowledgements}
M. Z. is supported by the National Natural Science Foundation of China (Grant No. 12005080) and Young Talents Foundation of Jiangxi Normal University (Grant No. 12020779). J. J.  is supported by the National Natural Science Foundation of China (Grants No. 11775022 and No. 11873044).


\begin{thebibliography}{58}%
\makeatletter
\providecommand \@ifxundefined [1]{%
 \@ifx{#1\undefined}
}%
\providecommand \@ifnum [1]{%
 \ifnum #1\expandafter \@firstoftwo
 \else \expandafter \@secondoftwo
 \fi
}%
\providecommand \@ifx [1]{%
 \ifx #1\expandafter \@firstoftwo
 \else \expandafter \@secondoftwo
 \fi
}%
\providecommand \natexlab [1]{#1}%
\providecommand \enquote  [1]{``#1''}%
\providecommand \bibnamefont  [1]{#1}%
\providecommand \bibfnamefont [1]{#1}%
\providecommand \citenamefont [1]{#1}%
\providecommand \href@noop [0]{\@secondoftwo}%
\providecommand \href [0]{\begingroup \@sanitize@url \@href}%
\providecommand \@href[1]{\@@startlink{#1}\@@href}%
\providecommand \@@href[1]{\endgroup#1\@@endlink}%
\providecommand \@sanitize@url [0]{\catcode `\\12\catcode `\$12\catcode
  `\&12\catcode `\#12\catcode `\^12\catcode `\_12\catcode `\%12\relax}%
\providecommand \@@startlink[1]{}%
\providecommand \@@endlink[0]{}%
\providecommand \url  [0]{\begingroup\@sanitize@url \@url }%
\providecommand \@url [1]{\endgroup\@href {#1}{\urlprefix }}%
\providecommand \urlprefix  [0]{URL }%
\providecommand \Eprint [0]{\href }%
\providecommand \doibase [0]{http://dx.doi.org/}%
\providecommand \selectlanguage [0]{\@gobble}%
\providecommand \bibinfo  [0]{\@secondoftwo}%
\providecommand \bibfield  [0]{\@secondoftwo}%
\providecommand \translation [1]{[#1]}%
\providecommand \BibitemOpen [0]{}%
\providecommand \bibitemStop [0]{}%
\providecommand \bibitemNoStop [0]{.\EOS\space}%
\providecommand \EOS [0]{\spacefactor3000\relax}%
\providecommand \BibitemShut  [1]{\csname bibitem#1\endcsname}%
\let\auto@bib@innerbib\@empty
%</preamble>
\bibitem [{\citenamefont {Cunha}\ \emph {et~al.}(2015)\citenamefont {Cunha},
  \citenamefont {Herdeiro}, \citenamefont {Radu},\ and\ \citenamefont
  {Runarsson}}]{Cunha:2015yba}%
  \BibitemOpen
  \bibfield  {author} {\bibinfo {author} {\bibfnamefont {P.~V.~P.}\
  \bibnamefont {Cunha}}, \bibinfo {author} {\bibfnamefont {C.~A.~R.}\
  \bibnamefont {Herdeiro}}, \bibinfo {author} {\bibfnamefont {E.}~\bibnamefont
  {Radu}}, \ and\ \bibinfo {author} {\bibfnamefont {H.~F.}\ \bibnamefont
  {Runarsson}},\ }\href {\doibase 10.1103/PhysRevLett.115.211102} {\bibfield
  {journal} {\bibinfo  {journal} {Phys. Rev. Lett.}\ }\textbf {\bibinfo
  {volume} {115}},\ \bibinfo {pages} {211102} (\bibinfo {year} {2015})},\
  \Eprint {http://arxiv.org/abs/1509.00021} {arXiv:1509.00021 [gr-qc]}
  \BibitemShut {NoStop}%
\bibitem [{\citenamefont {Synge}(1966)}]{synge1966escape}%
  \BibitemOpen
  \bibfield  {author} {\bibinfo {author} {\bibfnamefont {J.}~\bibnamefont
  {Synge}},\ }\href@noop {} {\bibfield  {journal} {\bibinfo  {journal} {Monthly
  Notices of the Royal Astronomical Society}\ }\textbf {\bibinfo {volume}
  {131}},\ \bibinfo {pages} {463} (\bibinfo {year} {1966})}\BibitemShut
  {NoStop}%
\bibitem [{\citenamefont {Luminet}(1979)}]{luminet1979image}%
  \BibitemOpen
  \bibfield  {author} {\bibinfo {author} {\bibfnamefont {J.-P.}\ \bibnamefont
  {Luminet}},\ }\href@noop {} {\bibfield  {journal} {\bibinfo  {journal}
  {Astronomy and Astrophysics}\ }\textbf {\bibinfo {volume} {75}},\ \bibinfo
  {pages} {228} (\bibinfo {year} {1979})}\BibitemShut {NoStop}%
\bibitem [{\citenamefont {Hawking}\ \emph {et~al.}(1973)\citenamefont
  {Hawking}, \citenamefont {Carter}, \citenamefont {Bardeen}, \citenamefont
  {Gursky}, \citenamefont {Thorne}, \citenamefont {Ruffini}, \citenamefont
  {Novikov} \emph {et~al.}}]{hawking1973black}%
  \BibitemOpen
  \bibfield  {author} {\bibinfo {author} {\bibfnamefont {S.}~\bibnamefont
  {Hawking}}, \bibinfo {author} {\bibfnamefont {B.}~\bibnamefont {Carter}},
  \bibinfo {author} {\bibfnamefont {J.~M.}\ \bibnamefont {Bardeen}}, \bibinfo
  {author} {\bibfnamefont {H.}~\bibnamefont {Gursky}}, \bibinfo {author}
  {\bibfnamefont {K.~S.}\ \bibnamefont {Thorne}}, \bibinfo {author}
  {\bibfnamefont {R.}~\bibnamefont {Ruffini}}, \bibinfo {author} {\bibfnamefont
  {I.~D.}\ \bibnamefont {Novikov}},  \emph {et~al.},\ }\href@noop {} {\emph
  {\bibinfo {title} {Black Holes}}},\ Vol.~\bibinfo {volume} {23}\ (\bibinfo
  {publisher} {CRC Press},\ \bibinfo {year} {1973})\BibitemShut {NoStop}%
\bibitem [{\citenamefont {Gralla}\ \emph {et~al.}(2019)\citenamefont {Gralla},
  \citenamefont {Holz},\ and\ \citenamefont {Wald}}]{Gralla:2019xty}%
  \BibitemOpen
  \bibfield  {author} {\bibinfo {author} {\bibfnamefont {S.~E.}\ \bibnamefont
  {Gralla}}, \bibinfo {author} {\bibfnamefont {D.~E.}\ \bibnamefont {Holz}}, \
  and\ \bibinfo {author} {\bibfnamefont {R.~M.}\ \bibnamefont {Wald}},\ }\href
  {\doibase 10.1103/PhysRevD.100.024018} {\bibfield  {journal} {\bibinfo
  {journal} {Phys. Rev. D}\ }\textbf {\bibinfo {volume} {100}},\ \bibinfo
  {pages} {024018} (\bibinfo {year} {2019})},\ \Eprint
  {http://arxiv.org/abs/1906.00873} {arXiv:1906.00873 [astro-ph.HE]}
  \BibitemShut {NoStop}%
\bibitem [{\citenamefont {Hioki}\ and\ \citenamefont
  {Miyamoto}(2008)}]{Hioki:2008zw}%
  \BibitemOpen
  \bibfield  {author} {\bibinfo {author} {\bibfnamefont {K.}~\bibnamefont
  {Hioki}}\ and\ \bibinfo {author} {\bibfnamefont {U.}~\bibnamefont
  {Miyamoto}},\ }\href {\doibase 10.1103/PhysRevD.78.044007} {\bibfield
  {journal} {\bibinfo  {journal} {Phys.\ Rev.\ D}\ }\textbf {\bibinfo {volume}
  {78}},\ \bibinfo {pages} {044007} (\bibinfo {year} {2008})},\ \Eprint
  {http://arxiv.org/abs/0805.3146} {arXiv:0805.3146 [gr-qc]} \BibitemShut
  {NoStop}%
\bibitem [{\citenamefont {Grenzebach}\ \emph {et~al.}(2014)\citenamefont
  {Grenzebach}, \citenamefont {Perlick},\ and\ \citenamefont
  {L\"ammerzahl}}]{Grenzebach:2014fha}%
  \BibitemOpen
  \bibfield  {author} {\bibinfo {author} {\bibfnamefont {A.}~\bibnamefont
  {Grenzebach}}, \bibinfo {author} {\bibfnamefont {V.}~\bibnamefont {Perlick}},
  \ and\ \bibinfo {author} {\bibfnamefont {C.}~\bibnamefont {L\"ammerzahl}},\
  }\href {\doibase 10.1103/PhysRevD.89.124004} {\bibfield  {journal} {\bibinfo
  {journal} {Phys. Rev. D}\ }\textbf {\bibinfo {volume} {89}},\ \bibinfo
  {pages} {124004} (\bibinfo {year} {2014})},\ \Eprint
  {http://arxiv.org/abs/1403.5234} {arXiv:1403.5234 [gr-qc]} \BibitemShut
  {NoStop}%
\bibitem [{\citenamefont {Wang}\ \emph {et~al.}(2018)\citenamefont {Wang},
  \citenamefont {Chen},\ and\ \citenamefont {Jing}}]{Wang:2017qhh}%
  \BibitemOpen
  \bibfield  {author} {\bibinfo {author} {\bibfnamefont {M.}~\bibnamefont
  {Wang}}, \bibinfo {author} {\bibfnamefont {S.}~\bibnamefont {Chen}}, \ and\
  \bibinfo {author} {\bibfnamefont {J.}~\bibnamefont {Jing}},\ }\href {\doibase
  10.1103/PhysRevD.97.064029} {\bibfield  {journal} {\bibinfo  {journal}
  {Phys.\ Rev.\ D}\ }\textbf {\bibinfo {volume} {97}},\ \bibinfo {pages}
  {064029} (\bibinfo {year} {2018})},\ \Eprint
  {http://arxiv.org/abs/1710.07172} {arXiv:1710.07172 [gr-qc]} \BibitemShut
  {NoStop}%
\bibitem [{\citenamefont {Guo}\ \emph {et~al.}(2018)\citenamefont {Guo},
  \citenamefont {Obers},\ and\ \citenamefont {Yan}}]{Guo:2018kis}%
  \BibitemOpen
  \bibfield  {author} {\bibinfo {author} {\bibfnamefont {M.}~\bibnamefont
  {Guo}}, \bibinfo {author} {\bibfnamefont {N.~A.}\ \bibnamefont {Obers}}, \
  and\ \bibinfo {author} {\bibfnamefont {H.}~\bibnamefont {Yan}},\ }\href
  {\doibase 10.1103/PhysRevD.98.084063} {\bibfield  {journal} {\bibinfo
  {journal} {Phys.\ Rev.\ D}\ }\textbf {\bibinfo {volume} {98}},\ \bibinfo
  {pages} {084063} (\bibinfo {year} {2018})},\ \Eprint
  {http://arxiv.org/abs/1806.05249} {arXiv:1806.05249 [gr-qc]} \BibitemShut
  {NoStop}%
\bibitem [{\citenamefont {Hennigar}\ \emph {et~al.}(2018)\citenamefont
  {Hennigar}, \citenamefont {Poshteh},\ and\ \citenamefont
  {Mann}}]{Hennigar:2018hza}%
  \BibitemOpen
  \bibfield  {author} {\bibinfo {author} {\bibfnamefont {R.~A.}\ \bibnamefont
  {Hennigar}}, \bibinfo {author} {\bibfnamefont {M.~B.~J.}\ \bibnamefont
  {Poshteh}}, \ and\ \bibinfo {author} {\bibfnamefont {R.~B.}\ \bibnamefont
  {Mann}},\ }\href {\doibase 10.1103/PhysRevD.97.064041} {\bibfield  {journal}
  {\bibinfo  {journal} {Phys.\ Rev.\ D}\ }\textbf {\bibinfo {volume} {97}},\
  \bibinfo {pages} {064041} (\bibinfo {year} {2018})},\ \Eprint
  {http://arxiv.org/abs/1801.03223} {arXiv:1801.03223 [gr-qc]} \BibitemShut
  {NoStop}%
\bibitem [{\citenamefont {Konoplya}(2019)}]{Konoplya:2019sns}%
  \BibitemOpen
  \bibfield  {author} {\bibinfo {author} {\bibfnamefont {R.}~\bibnamefont
  {Konoplya}},\ }\href {\doibase 10.1016/j.physletb.2019.05.043} {\bibfield
  {journal} {\bibinfo  {journal} {Phys.\ Lett.\ B}\ }\textbf {\bibinfo {volume}
  {795}},\ \bibinfo {pages} {1} (\bibinfo {year} {2019})},\ \Eprint
  {http://arxiv.org/abs/1905.00064} {arXiv:1905.00064 [gr-qc]} \BibitemShut
  {NoStop}%
\bibitem [{\citenamefont {Bambi}\ and\ \citenamefont
  {Yoshida}(2010)}]{Bambi:2010hf}%
  \BibitemOpen
  \bibfield  {author} {\bibinfo {author} {\bibfnamefont {C.}~\bibnamefont
  {Bambi}}\ and\ \bibinfo {author} {\bibfnamefont {N.}~\bibnamefont
  {Yoshida}},\ }\href {\doibase 10.1088/0264-9381/27/20/205006} {\bibfield
  {journal} {\bibinfo  {journal} {Class.\ Quant.\ Grav.}\ }\textbf {\bibinfo
  {volume} {27}},\ \bibinfo {pages} {205006} (\bibinfo {year} {2010})},\
  \Eprint {http://arxiv.org/abs/1004.3149} {arXiv:1004.3149 [gr-qc]}
  \BibitemShut {NoStop}%
\bibitem [{\citenamefont {Konoplya}\ \emph {et~al.}(2020)\citenamefont
  {Konoplya}, \citenamefont {Pappas},\ and\ \citenamefont
  {Zhidenko}}]{Konoplya:2019fpy}%
  \BibitemOpen
  \bibfield  {author} {\bibinfo {author} {\bibfnamefont {R.~A.}\ \bibnamefont
  {Konoplya}}, \bibinfo {author} {\bibfnamefont {T.}~\bibnamefont {Pappas}}, \
  and\ \bibinfo {author} {\bibfnamefont {A.}~\bibnamefont {Zhidenko}},\ }\href
  {\doibase 10.1103/PhysRevD.101.044054} {\bibfield  {journal} {\bibinfo
  {journal} {Phys.\ Rev.\ D}\ }\textbf {\bibinfo {volume} {101}},\ \bibinfo
  {pages} {044054} (\bibinfo {year} {2020})},\ \Eprint
  {http://arxiv.org/abs/1907.10112} {arXiv:1907.10112 [gr-qc]} \BibitemShut
  {NoStop}%
\bibitem [{\citenamefont {Wei}\ \emph {et~al.}(2019)\citenamefont {Wei},
  \citenamefont {Liu},\ and\ \citenamefont {Mann}}]{Wei:2018xks}%
  \BibitemOpen
  \bibfield  {author} {\bibinfo {author} {\bibfnamefont {S.-W.}\ \bibnamefont
  {Wei}}, \bibinfo {author} {\bibfnamefont {Y.-X.}\ \bibnamefont {Liu}}, \ and\
  \bibinfo {author} {\bibfnamefont {R.~B.}\ \bibnamefont {Mann}},\ }\href
  {\doibase 10.1103/PhysRevD.99.041303} {\bibfield  {journal} {\bibinfo
  {journal} {Phys. Rev. D}\ }\textbf {\bibinfo {volume} {99}},\ \bibinfo
  {pages} {041303} (\bibinfo {year} {2019})},\ \Eprint
  {http://arxiv.org/abs/1811.00047} {arXiv:1811.00047 [gr-qc]} \BibitemShut
  {NoStop}%
\bibitem [{\citenamefont {Jusufi}\ \emph {et~al.}(2021)\citenamefont {Jusufi},
  \citenamefont {Azreg-A\"\i{}nou}, \citenamefont {Jamil}, \citenamefont {Wei},
  \citenamefont {Wu},\ and\ \citenamefont {Wang}}]{Jusufi:2020odz}%
  \BibitemOpen
  \bibfield  {author} {\bibinfo {author} {\bibfnamefont {K.}~\bibnamefont
  {Jusufi}}, \bibinfo {author} {\bibfnamefont {M.}~\bibnamefont
  {Azreg-A\"\i{}nou}}, \bibinfo {author} {\bibfnamefont {M.}~\bibnamefont
  {Jamil}}, \bibinfo {author} {\bibfnamefont {S.-W.}\ \bibnamefont {Wei}},
  \bibinfo {author} {\bibfnamefont {Q.}~\bibnamefont {Wu}}, \ and\ \bibinfo
  {author} {\bibfnamefont {A.}~\bibnamefont {Wang}},\ }\href {\doibase
  10.1103/PhysRevD.103.024013} {\bibfield  {journal} {\bibinfo  {journal}
  {Phys. Rev. D}\ }\textbf {\bibinfo {volume} {103}},\ \bibinfo {pages}
  {024013} (\bibinfo {year} {2021})},\ \Eprint
  {http://arxiv.org/abs/2008.08450} {arXiv:2008.08450 [gr-qc]} \BibitemShut
  {NoStop}%
\bibitem [{\citenamefont {Wang}\ \emph {et~al.}(2019)\citenamefont {Wang},
  \citenamefont {Xu},\ and\ \citenamefont {Wei}}]{Wang:2018prk}%
  \BibitemOpen
  \bibfield  {author} {\bibinfo {author} {\bibfnamefont {H.-M.}\ \bibnamefont
  {Wang}}, \bibinfo {author} {\bibfnamefont {Y.-M.}\ \bibnamefont {Xu}}, \ and\
  \bibinfo {author} {\bibfnamefont {S.-W.}\ \bibnamefont {Wei}},\ }\href
  {\doibase 10.1088/1475-7516/2019/03/046} {\bibfield  {journal} {\bibinfo
  {journal} {JCAP}\ }\textbf {\bibinfo {volume} {03}},\ \bibinfo {pages} {046}
  (\bibinfo {year} {2019})},\ \Eprint {http://arxiv.org/abs/1810.12767}
  {arXiv:1810.12767 [gr-qc]} \BibitemShut {NoStop}%
\bibitem [{\citenamefont {Liu}\ \emph {et~al.}(2020)\citenamefont {Liu},
  \citenamefont {Zhu}, \citenamefont {Wu}, \citenamefont {Jusufi},
  \citenamefont {Jamil}, \citenamefont {Azreg-A\"\i{}nou},\ and\ \citenamefont
  {Wang}}]{Liu:2020ola}%
  \BibitemOpen
  \bibfield  {author} {\bibinfo {author} {\bibfnamefont {C.}~\bibnamefont
  {Liu}}, \bibinfo {author} {\bibfnamefont {T.}~\bibnamefont {Zhu}}, \bibinfo
  {author} {\bibfnamefont {Q.}~\bibnamefont {Wu}}, \bibinfo {author}
  {\bibfnamefont {K.}~\bibnamefont {Jusufi}}, \bibinfo {author} {\bibfnamefont
  {M.}~\bibnamefont {Jamil}}, \bibinfo {author} {\bibfnamefont
  {M.}~\bibnamefont {Azreg-A\"\i{}nou}}, \ and\ \bibinfo {author}
  {\bibfnamefont {A.}~\bibnamefont {Wang}},\ }\href {\doibase
  10.1103/PhysRevD.101.084001} {\bibfield  {journal} {\bibinfo  {journal}
  {Phys. Rev. D}\ }\textbf {\bibinfo {volume} {101}},\ \bibinfo {pages}
  {084001} (\bibinfo {year} {2020})},\ \Eprint
  {http://arxiv.org/abs/2003.00477} {arXiv:2003.00477 [gr-qc]} \BibitemShut
  {NoStop}%
\bibitem [{\citenamefont {Kumar}\ \emph {et~al.}(2019)\citenamefont {Kumar},
  \citenamefont {Ghosh},\ and\ \citenamefont {Wang}}]{Kumar:2019pjp}%
  \BibitemOpen
  \bibfield  {author} {\bibinfo {author} {\bibfnamefont {R.}~\bibnamefont
  {Kumar}}, \bibinfo {author} {\bibfnamefont {S.~G.}\ \bibnamefont {Ghosh}}, \
  and\ \bibinfo {author} {\bibfnamefont {A.}~\bibnamefont {Wang}},\ }\href
  {\doibase 10.1103/PhysRevD.100.124024} {\bibfield  {journal} {\bibinfo
  {journal} {Phys. Rev. D}\ }\textbf {\bibinfo {volume} {100}},\ \bibinfo
  {pages} {124024} (\bibinfo {year} {2019})},\ \Eprint
  {http://arxiv.org/abs/1912.05154} {arXiv:1912.05154 [gr-qc]} \BibitemShut
  {NoStop}%
\bibitem [{\citenamefont {Chang}\ and\ \citenamefont
  {Zhu}(2020)}]{Chang:2020miq}%
  \BibitemOpen
  \bibfield  {author} {\bibinfo {author} {\bibfnamefont {Z.}~\bibnamefont
  {Chang}}\ and\ \bibinfo {author} {\bibfnamefont {Q.-H.}\ \bibnamefont
  {Zhu}},\ }\href {\doibase 10.1103/PhysRevD.101.084029} {\bibfield  {journal}
  {\bibinfo  {journal} {Phys. Rev. D}\ }\textbf {\bibinfo {volume} {101}},\
  \bibinfo {pages} {084029} (\bibinfo {year} {2020})},\ \Eprint
  {http://arxiv.org/abs/2001.05175} {arXiv:2001.05175 [gr-qc]} \BibitemShut
  {NoStop}%
\bibitem [{\citenamefont {Zeng}\ \emph {et~al.}(2020)\citenamefont {Zeng},
  \citenamefont {Zhang},\ and\ \citenamefont {Zhang}}]{Zeng:2020dco}%
  \BibitemOpen
  \bibfield  {author} {\bibinfo {author} {\bibfnamefont {X.-X.}\ \bibnamefont
  {Zeng}}, \bibinfo {author} {\bibfnamefont {H.-Q.}\ \bibnamefont {Zhang}}, \
  and\ \bibinfo {author} {\bibfnamefont {H.}~\bibnamefont {Zhang}},\ }\href
  {\doibase 10.1140/epjc/s10052-020-08449-y} {\bibfield  {journal} {\bibinfo
  {journal} {Eur. Phys. J. C}\ }\textbf {\bibinfo {volume} {80}},\ \bibinfo
  {pages} {872} (\bibinfo {year} {2020})},\ \Eprint
  {http://arxiv.org/abs/2004.12074} {arXiv:2004.12074 [gr-qc]} \BibitemShut
  {NoStop}%
\bibitem [{\citenamefont {Zeng}\ and\ \citenamefont
  {Zhang}(2020)}]{Zeng:2020vsj}%
  \BibitemOpen
  \bibfield  {author} {\bibinfo {author} {\bibfnamefont {X.-X.}\ \bibnamefont
  {Zeng}}\ and\ \bibinfo {author} {\bibfnamefont {H.-Q.}\ \bibnamefont
  {Zhang}},\ }\href {\doibase 10.1140/epjc/s10052-020-08656-7} {\bibfield
  {journal} {\bibinfo  {journal} {Eur. Phys. J. C}\ }\textbf {\bibinfo {volume}
  {80}},\ \bibinfo {pages} {1058} (\bibinfo {year} {2020})},\ \Eprint
  {http://arxiv.org/abs/2007.06333} {arXiv:2007.06333 [gr-qc]} \BibitemShut
  {NoStop}%
\bibitem [{\citenamefont {Xavier}\ \emph {et~al.}(2020)\citenamefont {Xavier},
  \citenamefont {Cunha}, \citenamefont {Crispino},\ and\ \citenamefont
  {Herdeiro}}]{Xavier:2020egv}%
  \BibitemOpen
  \bibfield  {author} {\bibinfo {author} {\bibfnamefont {S.~V.~M.}\
  \bibnamefont {Xavier}}, \bibinfo {author} {\bibfnamefont {P.~V.}\
  \bibnamefont {Cunha}}, \bibinfo {author} {\bibfnamefont {L.~C.}\ \bibnamefont
  {Crispino}}, \ and\ \bibinfo {author} {\bibfnamefont {C.~A.}\ \bibnamefont
  {Herdeiro}},\ }\href {\doibase 10.1142/S0218271820410059} {\bibfield
  {journal} {\bibinfo  {journal} {Int. J. Mod. Phys. D}\ }\textbf {\bibinfo
  {volume} {29}},\ \bibinfo {pages} {2041005} (\bibinfo {year} {2020})},\
  \Eprint {http://arxiv.org/abs/2003.14349} {arXiv:2003.14349 [gr-qc]}
  \BibitemShut {NoStop}%
\bibitem [{\citenamefont {Hu}\ \emph {et~al.}(2021)\citenamefont {Hu},
  \citenamefont {Zhong}, \citenamefont {Li}, \citenamefont {Guo},\ and\
  \citenamefont {Chen}}]{Hu:2020usx}%
  \BibitemOpen
  \bibfield  {author} {\bibinfo {author} {\bibfnamefont {Z.}~\bibnamefont
  {Hu}}, \bibinfo {author} {\bibfnamefont {Z.}~\bibnamefont {Zhong}}, \bibinfo
  {author} {\bibfnamefont {P.-C.}\ \bibnamefont {Li}}, \bibinfo {author}
  {\bibfnamefont {M.}~\bibnamefont {Guo}}, \ and\ \bibinfo {author}
  {\bibfnamefont {B.}~\bibnamefont {Chen}},\ }\href {\doibase
  10.1103/PhysRevD.103.044057} {\bibfield  {journal} {\bibinfo  {journal}
  {Phys. Rev. D}\ }\textbf {\bibinfo {volume} {103}},\ \bibinfo {pages}
  {044057} (\bibinfo {year} {2021})},\ \Eprint
  {http://arxiv.org/abs/2012.07022} {arXiv:2012.07022 [gr-qc]} \BibitemShut
  {NoStop}%
\bibitem [{\citenamefont {Jusufi}(2020)}]{Jusufi:2019ltj}%
  \BibitemOpen
  \bibfield  {author} {\bibinfo {author} {\bibfnamefont {K.}~\bibnamefont
  {Jusufi}},\ }\href {\doibase 10.1103/PhysRevD.101.084055} {\bibfield
  {journal} {\bibinfo  {journal} {Phys. Rev. D}\ }\textbf {\bibinfo {volume}
  {101}},\ \bibinfo {pages} {084055} (\bibinfo {year} {2020})},\ \Eprint
  {http://arxiv.org/abs/1912.13320} {arXiv:1912.13320 [gr-qc]} \BibitemShut
  {NoStop}%
\bibitem [{\citenamefont {Ghasemi-Nodehi}\ \emph {et~al.}(2020)\citenamefont
  {Ghasemi-Nodehi}, \citenamefont {Azreg-A\"\i{}nou}, \citenamefont {Jusufi},\
  and\ \citenamefont {Jamil}}]{Ghasemi-Nodehi:2020oiz}%
  \BibitemOpen
  \bibfield  {author} {\bibinfo {author} {\bibfnamefont {M.}~\bibnamefont
  {Ghasemi-Nodehi}}, \bibinfo {author} {\bibfnamefont {M.}~\bibnamefont
  {Azreg-A\"\i{}nou}}, \bibinfo {author} {\bibfnamefont {K.}~\bibnamefont
  {Jusufi}}, \ and\ \bibinfo {author} {\bibfnamefont {M.}~\bibnamefont
  {Jamil}},\ }\href {\doibase 10.1103/PhysRevD.102.104032} {\bibfield
  {journal} {\bibinfo  {journal} {Phys. Rev. D}\ }\textbf {\bibinfo {volume}
  {102}},\ \bibinfo {pages} {104032} (\bibinfo {year} {2020})},\ \Eprint
  {http://arxiv.org/abs/2011.02276} {arXiv:2011.02276 [gr-qc]} \BibitemShut
  {NoStop}%
\bibitem [{\citenamefont {Lu}\ and\ \citenamefont {Lyu}(2020)}]{Lu:2019zxb}%
  \BibitemOpen
  \bibfield  {author} {\bibinfo {author} {\bibfnamefont {H.}~\bibnamefont
  {Lu}}\ and\ \bibinfo {author} {\bibfnamefont {H.-D.}\ \bibnamefont {Lyu}},\
  }\href {\doibase 10.1103/PhysRevD.101.044059} {\bibfield  {journal} {\bibinfo
   {journal} {Phys. Rev. D}\ }\textbf {\bibinfo {volume} {101}},\ \bibinfo
  {pages} {044059} (\bibinfo {year} {2020})},\ \Eprint
  {http://arxiv.org/abs/1911.02019} {arXiv:1911.02019 [gr-qc]} \BibitemShut
  {NoStop}%
\bibitem [{\citenamefont {Feng}\ and\ \citenamefont {Lu}(2020)}]{Feng:2019zzn}%
  \BibitemOpen
  \bibfield  {author} {\bibinfo {author} {\bibfnamefont {X.-H.}\ \bibnamefont
  {Feng}}\ and\ \bibinfo {author} {\bibfnamefont {H.}~\bibnamefont {Lu}},\
  }\href {\doibase 10.1140/epjc/s10052-020-8119-z} {\bibfield  {journal}
  {\bibinfo  {journal} {Eur. Phys. J. C}\ }\textbf {\bibinfo {volume} {80}},\
  \bibinfo {pages} {551} (\bibinfo {year} {2020})},\ \Eprint
  {http://arxiv.org/abs/1911.12368} {arXiv:1911.12368 [gr-qc]} \BibitemShut
  {NoStop}%
\bibitem [{\citenamefont {Hendi}\ and\ \citenamefont
  {Jafarzade}(2020)}]{Hendi:2020ebh}%
  \BibitemOpen
  \bibfield  {author} {\bibinfo {author} {\bibfnamefont {S.~H.}\ \bibnamefont
  {Hendi}}\ and\ \bibinfo {author} {\bibfnamefont {K.}~\bibnamefont
  {Jafarzade}},\ }\href@noop {} {\  (\bibinfo {year} {2020})},\ \Eprint
  {http://arxiv.org/abs/2012.13271} {arXiv:2012.13271 [hep-th]} \BibitemShut
  {NoStop}%
\bibitem [{\citenamefont {Cunha}\ \emph
  {et~al.}(2018{\natexlab{a}})\citenamefont {Cunha}, \citenamefont {Herdeiro},\
  and\ \citenamefont {Rodriguez}}]{Cunha:2018cof}%
  \BibitemOpen
  \bibfield  {author} {\bibinfo {author} {\bibfnamefont {P.~V.~P.}\
  \bibnamefont {Cunha}}, \bibinfo {author} {\bibfnamefont {C.~A.~R.}\
  \bibnamefont {Herdeiro}}, \ and\ \bibinfo {author} {\bibfnamefont {M.~J.}\
  \bibnamefont {Rodriguez}},\ }\href {\doibase 10.1103/PhysRevD.98.044053}
  {\bibfield  {journal} {\bibinfo  {journal} {Phys. Rev. D}\ }\textbf {\bibinfo
  {volume} {98}},\ \bibinfo {pages} {044053} (\bibinfo {year}
  {2018}{\natexlab{a}})},\ \Eprint {http://arxiv.org/abs/1805.03798}
  {arXiv:1805.03798 [gr-qc]} \BibitemShut {NoStop}%
\bibitem [{\citenamefont {Cunha}\ \emph
  {et~al.}(2018{\natexlab{b}})\citenamefont {Cunha}, \citenamefont {Herdeiro},\
  and\ \citenamefont {Rodriguez}}]{Cunha:2018gql}%
  \BibitemOpen
  \bibfield  {author} {\bibinfo {author} {\bibfnamefont {P.~V.~P.}\
  \bibnamefont {Cunha}}, \bibinfo {author} {\bibfnamefont {C.~A.~R.}\
  \bibnamefont {Herdeiro}}, \ and\ \bibinfo {author} {\bibfnamefont {M.~J.}\
  \bibnamefont {Rodriguez}},\ }\href {\doibase 10.1103/PhysRevD.97.084020}
  {\bibfield  {journal} {\bibinfo  {journal} {Phys. Rev. D}\ }\textbf {\bibinfo
  {volume} {97}},\ \bibinfo {pages} {084020} (\bibinfo {year}
  {2018}{\natexlab{b}})},\ \Eprint {http://arxiv.org/abs/1802.02675}
  {arXiv:1802.02675 [gr-qc]} \BibitemShut {NoStop}%
\bibitem [{\citenamefont {Gralla}\ and\ \citenamefont
  {Lupsasca}(2020)}]{Gralla:2020yvo}%
  \BibitemOpen
  \bibfield  {author} {\bibinfo {author} {\bibfnamefont {S.~E.}\ \bibnamefont
  {Gralla}}\ and\ \bibinfo {author} {\bibfnamefont {A.}~\bibnamefont
  {Lupsasca}},\ }\href {\doibase 10.1103/PhysRevD.102.124003} {\bibfield
  {journal} {\bibinfo  {journal} {Phys. Rev. D}\ }\textbf {\bibinfo {volume}
  {102}},\ \bibinfo {pages} {124003} (\bibinfo {year} {2020})},\ \Eprint
  {http://arxiv.org/abs/2007.10336} {arXiv:2007.10336 [gr-qc]} \BibitemShut
  {NoStop}%
\bibitem [{\citenamefont {Johnson}\ \emph {et~al.}(2020)\citenamefont {Johnson}
  \emph {et~al.}}]{Johnson:2019ljv}%
  \BibitemOpen
  \bibfield  {author} {\bibinfo {author} {\bibfnamefont {M.~D.}\ \bibnamefont
  {Johnson}} \emph {et~al.},\ }\href {\doibase 10.1126/sciadv.aaz1310}
  {\bibfield  {journal} {\bibinfo  {journal} {Sci. Adv.}\ }\textbf {\bibinfo
  {volume} {6}},\ \bibinfo {pages} {eaaz1310} (\bibinfo {year} {2020})},\
  \Eprint {http://arxiv.org/abs/1907.04329} {arXiv:1907.04329 [astro-ph.IM]}
  \BibitemShut {NoStop}%
\bibitem [{\citenamefont {Zhang}\ and\ \citenamefont
  {Guo}(2020)}]{Zhang:2019glo}%
  \BibitemOpen
  \bibfield  {author} {\bibinfo {author} {\bibfnamefont {M.}~\bibnamefont
  {Zhang}}\ and\ \bibinfo {author} {\bibfnamefont {M.}~\bibnamefont {Guo}},\
  }\href {\doibase 10.1140/epjc/s10052-020-8389-5} {\bibfield  {journal}
  {\bibinfo  {journal} {Eur. Phys. J. C}\ }\textbf {\bibinfo {volume} {80}},\
  \bibinfo {pages} {790} (\bibinfo {year} {2020})},\ \Eprint
  {http://arxiv.org/abs/1909.07033} {arXiv:1909.07033 [gr-qc]} \BibitemShut
  {NoStop}%
\bibitem [{\citenamefont {Hioki}\ and\ \citenamefont
  {Maeda}(2009)}]{Hioki:2009na}%
  \BibitemOpen
  \bibfield  {author} {\bibinfo {author} {\bibfnamefont {K.}~\bibnamefont
  {Hioki}}\ and\ \bibinfo {author} {\bibfnamefont {K.-i.}\ \bibnamefont
  {Maeda}},\ }\href {\doibase 10.1103/PhysRevD.80.024042} {\bibfield  {journal}
  {\bibinfo  {journal} {Phys. Rev.}\ }\textbf {\bibinfo {volume} {D80}},\
  \bibinfo {pages} {024042} (\bibinfo {year} {2009})},\ \Eprint
  {http://arxiv.org/abs/0904.3575} {arXiv:0904.3575 [astro-ph.HE]} \BibitemShut
  {NoStop}%
%%CITATION = ARXIV:0904.3575;%%
\bibitem [{\citenamefont {Ohgami}\ and\ \citenamefont
  {Sakai}(2015)}]{Ohgami:2015nra}%
  \BibitemOpen
  \bibfield  {author} {\bibinfo {author} {\bibfnamefont {T.}~\bibnamefont
  {Ohgami}}\ and\ \bibinfo {author} {\bibfnamefont {N.}~\bibnamefont {Sakai}},\
  }\href {\doibase 10.1103/PhysRevD.91.124020} {\bibfield  {journal} {\bibinfo
  {journal} {Phys.\ Rev.\ D}\ }\textbf {\bibinfo {volume} {91}},\ \bibinfo
  {pages} {124020} (\bibinfo {year} {2015})},\ \Eprint
  {http://arxiv.org/abs/1704.07065} {arXiv:1704.07065 [gr-qc]} \BibitemShut
  {NoStop}%
\bibitem [{\citenamefont {Nedkova}\ \emph {et~al.}(2013)\citenamefont
  {Nedkova}, \citenamefont {Tinchev},\ and\ \citenamefont
  {Yazadjiev}}]{Nedkova:2013msa}%
  \BibitemOpen
  \bibfield  {author} {\bibinfo {author} {\bibfnamefont {P.~G.}\ \bibnamefont
  {Nedkova}}, \bibinfo {author} {\bibfnamefont {V.~K.}\ \bibnamefont
  {Tinchev}}, \ and\ \bibinfo {author} {\bibfnamefont {S.~S.}\ \bibnamefont
  {Yazadjiev}},\ }\href {\doibase 10.1103/PhysRevD.88.124019} {\bibfield
  {journal} {\bibinfo  {journal} {Phys.\ Rev.\ D}\ }\textbf {\bibinfo {volume}
  {88}},\ \bibinfo {pages} {124019} (\bibinfo {year} {2013})},\ \Eprint
  {http://arxiv.org/abs/1307.7647} {arXiv:1307.7647 [gr-qc]} \BibitemShut
  {NoStop}%
\bibitem [{\citenamefont {Shaikh}(2018)}]{Shaikh:2018kfv}%
  \BibitemOpen
  \bibfield  {author} {\bibinfo {author} {\bibfnamefont {R.}~\bibnamefont
  {Shaikh}},\ }\href {\doibase 10.1103/PhysRevD.98.024044} {\bibfield
  {journal} {\bibinfo  {journal} {Phys.\ Rev.\ D}\ }\textbf {\bibinfo {volume}
  {98}},\ \bibinfo {pages} {024044} (\bibinfo {year} {2018})},\ \Eprint
  {http://arxiv.org/abs/1803.11422} {arXiv:1803.11422 [gr-qc]} \BibitemShut
  {NoStop}%
\bibitem [{\citenamefont {Amir}\ \emph
  {et~al.}(2019{\natexlab{a}})\citenamefont {Amir}, \citenamefont {Banerjee},\
  and\ \citenamefont {Maharaj}}]{Amir:2018szm}%
  \BibitemOpen
  \bibfield  {author} {\bibinfo {author} {\bibfnamefont {M.}~\bibnamefont
  {Amir}}, \bibinfo {author} {\bibfnamefont {A.}~\bibnamefont {Banerjee}}, \
  and\ \bibinfo {author} {\bibfnamefont {S.~D.}\ \bibnamefont {Maharaj}},\
  }\href {\doibase 10.1016/j.aop.2018.11.004} {\bibfield  {journal} {\bibinfo
  {journal} {Annals Phys.}\ }\textbf {\bibinfo {volume} {400}},\ \bibinfo
  {pages} {198} (\bibinfo {year} {2019}{\natexlab{a}})},\ \Eprint
  {http://arxiv.org/abs/1805.12435} {arXiv:1805.12435 [gr-qc]} \BibitemShut
  {NoStop}%
\bibitem [{\citenamefont {Amir}\ \emph
  {et~al.}(2019{\natexlab{b}})\citenamefont {Amir}, \citenamefont {Jusufi},
  \citenamefont {Banerjee},\ and\ \citenamefont {Hansraj}}]{Amir:2018pcu}%
  \BibitemOpen
  \bibfield  {author} {\bibinfo {author} {\bibfnamefont {M.}~\bibnamefont
  {Amir}}, \bibinfo {author} {\bibfnamefont {K.}~\bibnamefont {Jusufi}},
  \bibinfo {author} {\bibfnamefont {A.}~\bibnamefont {Banerjee}}, \ and\
  \bibinfo {author} {\bibfnamefont {S.}~\bibnamefont {Hansraj}},\ }\href
  {\doibase 10.1088/1361-6382/ab42be} {\bibfield  {journal} {\bibinfo
  {journal} {Class.\ Quant.\ Grav.}\ }\textbf {\bibinfo {volume} {36}},\
  \bibinfo {pages} {215007} (\bibinfo {year} {2019}{\natexlab{b}})},\ \Eprint
  {http://arxiv.org/abs/1806.07782} {arXiv:1806.07782 [gr-qc]} \BibitemShut
  {NoStop}%
\bibitem [{\citenamefont {Wang}\ \emph {et~al.}(2020)\citenamefont {Wang},
  \citenamefont {Li}, \citenamefont {Zhang},\ and\ \citenamefont
  {Guo}}]{Wang:2020emr}%
  \BibitemOpen
  \bibfield  {author} {\bibinfo {author} {\bibfnamefont {X.}~\bibnamefont
  {Wang}}, \bibinfo {author} {\bibfnamefont {P.-C.}\ \bibnamefont {Li}},
  \bibinfo {author} {\bibfnamefont {C.-Y.}\ \bibnamefont {Zhang}}, \ and\
  \bibinfo {author} {\bibfnamefont {M.}~\bibnamefont {Guo}},\ }\href {\doibase
  10.1016/j.physletb.2020.135930} {\bibfield  {journal} {\bibinfo  {journal}
  {Phys. Lett. B}\ }\textbf {\bibinfo {volume} {811}},\ \bibinfo {pages}
  {135930} (\bibinfo {year} {2020})},\ \Eprint
  {http://arxiv.org/abs/2007.03327} {arXiv:2007.03327 [gr-qc]} \BibitemShut
  {NoStop}%
\bibitem [{\citenamefont {Johannsen}(2016)}]{Johannsen:2016uoh}%
  \BibitemOpen
  \bibfield  {author} {\bibinfo {author} {\bibfnamefont {T.}~\bibnamefont
  {Johannsen}},\ }\href {\doibase 10.1088/0264-9381/33/12/124001} {\bibfield
  {journal} {\bibinfo  {journal} {Class. Quant. Grav.}\ }\textbf {\bibinfo
  {volume} {33}},\ \bibinfo {pages} {124001} (\bibinfo {year} {2016})},\
  \Eprint {http://arxiv.org/abs/1602.07694} {arXiv:1602.07694 [astro-ph.HE]}
  \BibitemShut {NoStop}%
\bibitem [{\citenamefont {Akiyama}\ \emph {et~al.}(2019)\citenamefont {Akiyama}
  \emph {et~al.}}]{Akiyama:2019cqa}%
  \BibitemOpen
  \bibfield  {author} {\bibinfo {author} {\bibfnamefont {K.}~\bibnamefont
  {Akiyama}} \emph {et~al.} (\bibinfo {collaboration} {Event Horizon
  Telescope}),\ }\href {\doibase 10.3847/2041-8213/ab0ec7} {\bibfield
  {journal} {\bibinfo  {journal} {Astrophys.\ J.}\ }\textbf {\bibinfo {volume}
  {875}},\ \bibinfo {pages} {L1} (\bibinfo {year} {2019})},\ \Eprint
  {http://arxiv.org/abs/1906.11238} {arXiv:1906.11238 [astro-ph.GA]}
  \BibitemShut {NoStop}%
\bibitem [{\citenamefont {Zhang}\ and\ \citenamefont
  {Jiang}(2021)}]{Zhang:2020xub}%
  \BibitemOpen
  \bibfield  {author} {\bibinfo {author} {\bibfnamefont {M.}~\bibnamefont
  {Zhang}}\ and\ \bibinfo {author} {\bibfnamefont {J.}~\bibnamefont {Jiang}},\
  }\href {\doibase 10.1103/PhysRevD.103.025005} {\bibfield  {journal} {\bibinfo
   {journal} {Phys. Rev. D}\ }\textbf {\bibinfo {volume} {103}},\ \bibinfo
  {pages} {025005} (\bibinfo {year} {2021})},\ \Eprint
  {http://arxiv.org/abs/2010.12194} {arXiv:2010.12194 [gr-qc]} \BibitemShut
  {NoStop}%
\bibitem [{\citenamefont {Griffiths}\ and\ \citenamefont
  {Podolsky}(2006)}]{Griffiths:2005qp}%
  \BibitemOpen
  \bibfield  {author} {\bibinfo {author} {\bibfnamefont {J.~B.}\ \bibnamefont
  {Griffiths}}\ and\ \bibinfo {author} {\bibfnamefont {J.}~\bibnamefont
  {Podolsky}},\ }\href {\doibase 10.1142/S0218271806007742} {\bibfield
  {journal} {\bibinfo  {journal} {Int. J. Mod. Phys.}\ }\textbf {\bibinfo
  {volume} {D15}},\ \bibinfo {pages} {335} (\bibinfo {year} {2006})},\ \Eprint
  {http://arxiv.org/abs/gr-qc/0511091} {arXiv:gr-qc/0511091 [gr-qc]}
  \BibitemShut {NoStop}%
%%CITATION = GR-QC/0511091;%%
\bibitem [{\citenamefont {Abdujabbarov}\ \emph {et~al.}(2013)\citenamefont
  {Abdujabbarov}, \citenamefont {Atamurotov}, \citenamefont {Kucukakca},
  \citenamefont {Ahmedov},\ and\ \citenamefont {Camci}}]{Abdujabbarov:2012bn}%
  \BibitemOpen
  \bibfield  {author} {\bibinfo {author} {\bibfnamefont {A.}~\bibnamefont
  {Abdujabbarov}}, \bibinfo {author} {\bibfnamefont {F.}~\bibnamefont
  {Atamurotov}}, \bibinfo {author} {\bibfnamefont {Y.}~\bibnamefont
  {Kucukakca}}, \bibinfo {author} {\bibfnamefont {B.}~\bibnamefont {Ahmedov}},
  \ and\ \bibinfo {author} {\bibfnamefont {U.}~\bibnamefont {Camci}},\ }\href
  {\doibase 10.1007/s10509-012-1337-6} {\bibfield  {journal} {\bibinfo
  {journal} {Astrophys. Space Sci.}\ }\textbf {\bibinfo {volume} {344}},\
  \bibinfo {pages} {429} (\bibinfo {year} {2013})},\ \Eprint
  {http://arxiv.org/abs/1212.4949} {arXiv:1212.4949 [physics.gen-ph]}
  \BibitemShut {NoStop}%
\bibitem [{\citenamefont {Bardeen}(1972)}]{bardeen1972jm}%
  \BibitemOpen
  \bibfield  {author} {\bibinfo {author} {\bibfnamefont {J.~M.}\ \bibnamefont
  {Bardeen}},\ }\href@noop {} {\bibfield  {journal} {\bibinfo  {journal}
  {Astrophys. J.}\ }\textbf {\bibinfo {volume} {178}},\ \bibinfo {pages} {347}
  (\bibinfo {year} {1972})}\BibitemShut {NoStop}%
\bibitem [{\citenamefont {Manko}\ and\ \citenamefont
  {Ruiz}(2005)}]{Manko:2005nm}%
  \BibitemOpen
  \bibfield  {author} {\bibinfo {author} {\bibfnamefont {V.~S.}\ \bibnamefont
  {Manko}}\ and\ \bibinfo {author} {\bibfnamefont {E.}~\bibnamefont {Ruiz}},\
  }\href {\doibase 10.1088/0264-9381/22/17/014} {\bibfield  {journal} {\bibinfo
   {journal} {Class. Quant. Grav.}\ }\textbf {\bibinfo {volume} {22}},\
  \bibinfo {pages} {3555} (\bibinfo {year} {2005})},\ \Eprint
  {http://arxiv.org/abs/gr-qc/0505001} {arXiv:gr-qc/0505001} \BibitemShut
  {NoStop}%
\bibitem [{\citenamefont {Misner}(1963)}]{Misner:1963fr}%
  \BibitemOpen
  \bibfield  {author} {\bibinfo {author} {\bibfnamefont {C.~W.}\ \bibnamefont
  {Misner}},\ }\href {\doibase 10.1063/1.1704019} {\bibfield  {journal}
  {\bibinfo  {journal} {J. Math. Phys.}\ }\textbf {\bibinfo {volume} {4}},\
  \bibinfo {pages} {924} (\bibinfo {year} {1963})}\BibitemShut {NoStop}%
\bibitem [{\citenamefont {Cl\'ement}\ \emph {et~al.}(2015)\citenamefont
  {Cl\'ement}, \citenamefont {Gal'tsov},\ and\ \citenamefont
  {Guenouche}}]{Clement:2015cxa}%
  \BibitemOpen
  \bibfield  {author} {\bibinfo {author} {\bibfnamefont {G.}~\bibnamefont
  {Cl\'ement}}, \bibinfo {author} {\bibfnamefont {D.}~\bibnamefont {Gal'tsov}},
  \ and\ \bibinfo {author} {\bibfnamefont {M.}~\bibnamefont {Guenouche}},\
  }\href {\doibase 10.1016/j.physletb.2015.09.074} {\bibfield  {journal}
  {\bibinfo  {journal} {Phys. Lett. B}\ }\textbf {\bibinfo {volume} {750}},\
  \bibinfo {pages} {591} (\bibinfo {year} {2015})},\ \Eprint
  {http://arxiv.org/abs/1508.07622} {arXiv:1508.07622 [hep-th]} \BibitemShut
  {NoStop}%
\bibitem [{\citenamefont {Ballon~Bordo}\ \emph {et~al.}(2019)\citenamefont
  {Ballon~Bordo}, \citenamefont {Gray}, \citenamefont {Hennigar},\ and\
  \citenamefont {Kubiz\v{n}\'ak}}]{Bordo:2019rhu}%
  \BibitemOpen
  \bibfield  {author} {\bibinfo {author} {\bibfnamefont {A.}~\bibnamefont
  {Ballon~Bordo}}, \bibinfo {author} {\bibfnamefont {F.}~\bibnamefont {Gray}},
  \bibinfo {author} {\bibfnamefont {R.~A.}\ \bibnamefont {Hennigar}}, \ and\
  \bibinfo {author} {\bibfnamefont {D.}~\bibnamefont {Kubiz\v{n}\'ak}},\ }\href
  {\doibase 10.1016/j.physletb.2019.134972} {\bibfield  {journal} {\bibinfo
  {journal} {Phys. Lett. B}\ }\textbf {\bibinfo {volume} {798}},\ \bibinfo
  {pages} {134972} (\bibinfo {year} {2019})},\ \Eprint
  {http://arxiv.org/abs/1905.06350} {arXiv:1905.06350 [hep-th]} \BibitemShut
  {NoStop}%
\bibitem [{\citenamefont {Carter}(1968)}]{Carter:1968rr}%
  \BibitemOpen
  \bibfield  {author} {\bibinfo {author} {\bibfnamefont {B.}~\bibnamefont
  {Carter}},\ }\href {\doibase 10.1103/PhysRev.174.1559} {\bibfield  {journal}
  {\bibinfo  {journal} {Phys. Rev.}\ }\textbf {\bibinfo {volume} {174}},\
  \bibinfo {pages} {1559} (\bibinfo {year} {1968})}\BibitemShut {NoStop}%
\bibitem [{\citenamefont {Cebeci}\ \emph {et~al.}(2016)\citenamefont {Cebeci},
  \citenamefont {\"Ozdemir},\ and\ \citenamefont
  {\c{S}entorun}}]{Cebeci:2015fie}%
  \BibitemOpen
  \bibfield  {author} {\bibinfo {author} {\bibfnamefont {H.}~\bibnamefont
  {Cebeci}}, \bibinfo {author} {\bibfnamefont {N.}~\bibnamefont {\"Ozdemir}}, \
  and\ \bibinfo {author} {\bibfnamefont {S.}~\bibnamefont {\c{S}entorun}},\
  }\href {\doibase 10.1103/PhysRevD.93.104031} {\bibfield  {journal} {\bibinfo
  {journal} {Phys. Rev. D}\ }\textbf {\bibinfo {volume} {93}},\ \bibinfo
  {pages} {104031} (\bibinfo {year} {2016})},\ \Eprint
  {http://arxiv.org/abs/1512.08682} {arXiv:1512.08682 [gr-qc]} \BibitemShut
  {NoStop}%
\bibitem [{\citenamefont {Guo}\ and\ \citenamefont {Gao}(2020)}]{Guo:2020qwk}%
  \BibitemOpen
  \bibfield  {author} {\bibinfo {author} {\bibfnamefont {M.}~\bibnamefont
  {Guo}}\ and\ \bibinfo {author} {\bibfnamefont {S.}~\bibnamefont {Gao}},\
  }\href@noop {} {\  (\bibinfo {year} {2020})},\ \Eprint
  {http://arxiv.org/abs/2011.02211} {arXiv:2011.02211 [gr-qc]} \BibitemShut
  {NoStop}%
\bibitem [{\citenamefont {Cunha}\ \emph
  {et~al.}(2016{\natexlab{a}})\citenamefont {Cunha}, \citenamefont {Herdeiro},
  \citenamefont {Radu},\ and\ \citenamefont {Runarsson}}]{Cunha:2016bpi}%
  \BibitemOpen
  \bibfield  {author} {\bibinfo {author} {\bibfnamefont {P.~V.~P.}\
  \bibnamefont {Cunha}}, \bibinfo {author} {\bibfnamefont {C.~A.~R.}\
  \bibnamefont {Herdeiro}}, \bibinfo {author} {\bibfnamefont {E.}~\bibnamefont
  {Radu}}, \ and\ \bibinfo {author} {\bibfnamefont {H.~F.}\ \bibnamefont
  {Runarsson}},\ }\href {\doibase 10.1142/S0218271816410212} {\bibfield
  {journal} {\bibinfo  {journal} {Int. J. Mod. Phys. D}\ }\textbf {\bibinfo
  {volume} {25}},\ \bibinfo {pages} {1641021} (\bibinfo {year}
  {2016}{\natexlab{a}})},\ \Eprint {http://arxiv.org/abs/1605.08293}
  {arXiv:1605.08293 [gr-qc]} \BibitemShut {NoStop}%
\bibitem [{\citenamefont {Long}\ \emph {et~al.}(2019)\citenamefont {Long},
  \citenamefont {Chen}, \citenamefont {Wang},\ and\ \citenamefont
  {Jing}}]{Long:2018tij}%
  \BibitemOpen
  \bibfield  {author} {\bibinfo {author} {\bibfnamefont {F.}~\bibnamefont
  {Long}}, \bibinfo {author} {\bibfnamefont {S.}~\bibnamefont {Chen}}, \bibinfo
  {author} {\bibfnamefont {J.}~\bibnamefont {Wang}}, \ and\ \bibinfo {author}
  {\bibfnamefont {J.}~\bibnamefont {Jing}},\ }\href {\doibase
  10.1140/epjc/s10052-019-6989-8} {\bibfield  {journal} {\bibinfo  {journal}
  {Eur. Phys. J. C}\ }\textbf {\bibinfo {volume} {79}},\ \bibinfo {pages} {466}
  (\bibinfo {year} {2019})},\ \Eprint {http://arxiv.org/abs/1812.11463}
  {arXiv:1812.11463 [gr-qc]} \BibitemShut {NoStop}%
\bibitem [{\citenamefont {Cunha}\ \emph {et~al.}(2019)\citenamefont {Cunha},
  \citenamefont {Herdeiro},\ and\ \citenamefont {Radu}}]{Cunha:2019dwb}%
  \BibitemOpen
  \bibfield  {author} {\bibinfo {author} {\bibfnamefont {P.~V.~P.}\
  \bibnamefont {Cunha}}, \bibinfo {author} {\bibfnamefont {C.~A.~R.}\
  \bibnamefont {Herdeiro}}, \ and\ \bibinfo {author} {\bibfnamefont
  {E.}~\bibnamefont {Radu}},\ }\href {\doibase 10.1103/PhysRevLett.123.011101}
  {\bibfield  {journal} {\bibinfo  {journal} {Phys. Rev. Lett.}\ }\textbf
  {\bibinfo {volume} {123}},\ \bibinfo {pages} {011101} (\bibinfo {year}
  {2019})},\ \Eprint {http://arxiv.org/abs/1904.09997} {arXiv:1904.09997
  [gr-qc]} \BibitemShut {NoStop}%
\bibitem [{\citenamefont {Herdeiro}\ \emph {et~al.}(2021)\citenamefont
  {Herdeiro}, \citenamefont {Radu}, \citenamefont {Silva}, \citenamefont
  {Sotiriou},\ and\ \citenamefont {Yunes}}]{Herdeiro:2020wei}%
  \BibitemOpen
  \bibfield  {author} {\bibinfo {author} {\bibfnamefont {C.~A.~R.}\
  \bibnamefont {Herdeiro}}, \bibinfo {author} {\bibfnamefont {E.}~\bibnamefont
  {Radu}}, \bibinfo {author} {\bibfnamefont {H.~O.}\ \bibnamefont {Silva}},
  \bibinfo {author} {\bibfnamefont {T.~P.}\ \bibnamefont {Sotiriou}}, \ and\
  \bibinfo {author} {\bibfnamefont {N.}~\bibnamefont {Yunes}},\ }\href
  {\doibase 10.1103/PhysRevLett.126.011103} {\bibfield  {journal} {\bibinfo
  {journal} {Phys. Rev. Lett.}\ }\textbf {\bibinfo {volume} {126}},\ \bibinfo
  {pages} {011103} (\bibinfo {year} {2021})},\ \Eprint
  {http://arxiv.org/abs/2009.03904} {arXiv:2009.03904 [gr-qc]} \BibitemShut
  {NoStop}%
\bibitem [{\citenamefont {Cunha}\ \emph
  {et~al.}(2016{\natexlab{b}})\citenamefont {Cunha}, \citenamefont {Grover},
  \citenamefont {Herdeiro}, \citenamefont {Radu}, \citenamefont {Runarsson},\
  and\ \citenamefont {Wittig}}]{Cunha:2016bjh}%
  \BibitemOpen
  \bibfield  {author} {\bibinfo {author} {\bibfnamefont {P.~V.~P.}\
  \bibnamefont {Cunha}}, \bibinfo {author} {\bibfnamefont {J.}~\bibnamefont
  {Grover}}, \bibinfo {author} {\bibfnamefont {C.}~\bibnamefont {Herdeiro}},
  \bibinfo {author} {\bibfnamefont {E.}~\bibnamefont {Radu}}, \bibinfo {author}
  {\bibfnamefont {H.}~\bibnamefont {Runarsson}}, \ and\ \bibinfo {author}
  {\bibfnamefont {A.}~\bibnamefont {Wittig}},\ }\href {\doibase
  10.1103/PhysRevD.94.104023} {\bibfield  {journal} {\bibinfo  {journal} {Phys.
  Rev. D}\ }\textbf {\bibinfo {volume} {94}},\ \bibinfo {pages} {104023}
  (\bibinfo {year} {2016}{\natexlab{b}})},\ \Eprint
  {http://arxiv.org/abs/1609.01340} {arXiv:1609.01340 [gr-qc]} \BibitemShut
  {NoStop}%
\end{thebibliography}
\end{document}